\documentclass{aa}

\usepackage{graphicx}
\usepackage{pjournal}
\usepackage{natbib}
\usepackage{amssymb}

\begin{document}

\title{Separation of anomalous and synchrotron emissions using WMAP polarization data}

 \author{M.-A. Miville-Desch\^enes\inst{1,2}\and N. Ysard\inst{1} \and A. Lavabre\inst{1} \and N. Ponthieu\inst{1} 
\and J. F. Mac\'{\i}as-P\'erez\inst{3} \and J.~Aumont\inst{3,4} \and J.P. Bernard\inst{4} }
 \institute{Institut d'Astrophysique Spatiale (CNRS), Universit\'e Paris-Sud, b\^at. 121, 91405, Orsay, France 
\and Canadian Institute for Theoretical Astrophysics, University of Toronto, 60 St. George Street, Toronto, ON, M5S~3H8, Canada
\and LPSC, Universit\'e Joseph Fourier Grenoble 1, CNRS/IN2P3, Institut National Polytechnique de Grenoble,
53, av. des Martyrs, 38026 Grenoble, France
\and Centre d'Etude Spatiale des Rayonnements, 9 av du Colonel Roche, 31028, Toulouse, France}

 \offprints{Marc-Antoine Miville-Desch\^enes}
 \mail{mamd@ias.u-psud.fr}
 \date{\today}

\titlerunning{Anomalous and synchrotron emissions using WMAP polarization data}
\authorrunning{Miville-Desch\^enes, M.-A. et al.}

\abstract
{
Apart from its contribution to cosmology, the WMAP data brings new information on the Galactic interstellar medium.
In particular the polarization data provide constraints on the Galactic magnetic field and 
the synchrotron emission, while the intensity data can be used to study the anomalous microwave
emission. 
}
{
The main goals of this study is to use the information from both WMAP intensity and polarization 
data to do a separation of the Galactic components, with a focus on the synchrotron and anomalous emissions.  
}
{
Our analysis is made at 23~GHz where the signal-to-noise ratio is the highest and the estimate of the CMB map is less critical.
Our estimate of the synchrotron intensity is based on an extrapolation of the Haslam 408~MHz data
with a spatially varying spectral index constrained by the WMAP 23~GHz polarization data and 
a bi-symmetrical spiral model of the galactic magnetic field with a turbulent part following a -5/3 power law spectrum.
}
{
The 23 GHz polarization data are found to be compatible with a magnetic field 
with a pitch angle $p=-8.5^\circ$ and an amplitude of the turbulent part of the magnetic field 0.57 times the local
value of the field, in agreement with what is found using rotation measures of pulsars and polarized extinction by dust. 
The synchrotron spectral index between 408~MHz and 23~GHz obtained from polarization data and our model
of the magnetic field has a mean value of $\beta_s=-3.00$ with a limited spatial variation with a 
standard deviation of 0.06. When thermal dust, free-free and synchrotron are removed from the WMAP intensity data,
the residual anomalous emission is highly correlated with thermal dust emission with a spectrum
in agreement with spinning dust models.
}
{
Considering a classical model of the large scale Galactic magnetic field,
we show that the polarization data of WMAP are in favor of a soft synchrotron intensity 
highly correlated with the 408~MHz data. Furthermore the combination of the WMAP polarization and intensity
data brings strong evidence for the presence of unpolarized spinning dust emission in the 20-60~GHz range.
In preparation for the Planck mission this joint analysis of polarization and intensity data opens new perspective 
on the study of the Galactic interstellar medium and on the component separation exercise.
}


\maketitle

\section{Introduction}

One of the greatest challenges of observing the Cosmic Microwave Background (CMB) 
in the 20-200~GHz range resides in the separation between the CMB and foreground emission. 
This task is facilitated by the fact that the intensity of the emission from the Galactic interstellar medium 
reaches a minimum in this range. 
On the other hand, even if the Galactic emission is weak, it is still stronger than
the CMB over a significant fraction of the sky.
In addition the identification of the cosmological signal is complicated by the fact 
that several interstellar emissions are superimposed in this frequency range: free-free, synchrotron, 
thermal dust. There are also strong evidence of an excess of emission in this range, discovered
by \cite{kogut1996,kogut1996a} in the COBE-DMR data. This emission,
is remarkably well correlated with thermal dust emission at 100~$\mu$m
with a spectrum which can be confused with that of free-free or synchrotron emissions.

\cite{leitch1997} showed that the correlation of this excess with 100~$\mu$m extends to sub-degree scales
in the North Celestial Loop (NCL) region. The lack of correlation of the anomalous emission
with H$\alpha$ measurements, 
also highlighted by \cite{de_oliveira-costa2002} using Tenerife and COBE-DMR data,
led \cite{leitch1997} to conclude that free-free emission can only account for the data 
if it comes from gas with $T_e>10^6$~K. 
\cite{draine1998} ruled out this interpretation 
based on energetic arguments and suggested rotational or spinning dust
as a plausible emission mechanism \cite[]{draine1998a}. 
This interpretation received strong support from dedicated ground observations 
\cite[e.g][]{watson2005,fernandez-cerezo2006}
and from the analysis of the WMAP data 
\cite[]{lagache2003b,finkbeiner2004,davies2006}.

The WMAP team proposed an alternative explanation. \cite{bennett2003} and \cite{hinshaw2007} argued
that high-energy electrons, responsible for the synchrotron emission at WMAP frequencies, 
are produced in star forming regions which are also bright in the infrared, 
suggesting a natural correlation between synchrotron and dust emission.
As the synchrotron emission is less correlated with
dust emission at lower frequencies (e.g. 408~MHz), 
the dust correlated emission observed in the WMAP data is explained as a change 
in the synchrotron spectral index which 
can only be attributed to a local change of the energy distribution of cosmic rays.
Spatial variations of the correlation between the magnetic field intensity and the gas density 
would not produce a frequency dependent behavior (see ~\S~\ref{section:synchrotron}). 
This interpretation has one important drawback:
the correlation between the 23~GHz WMAP emission and the 100~$\mu$m emission
extends down to low column density cirrus clouds, 
where there is no star forming activities and no local production of cosmic rays
(see \cite{davies2006}).

\begin{figure}
\includegraphics[width=\linewidth, draft=false]{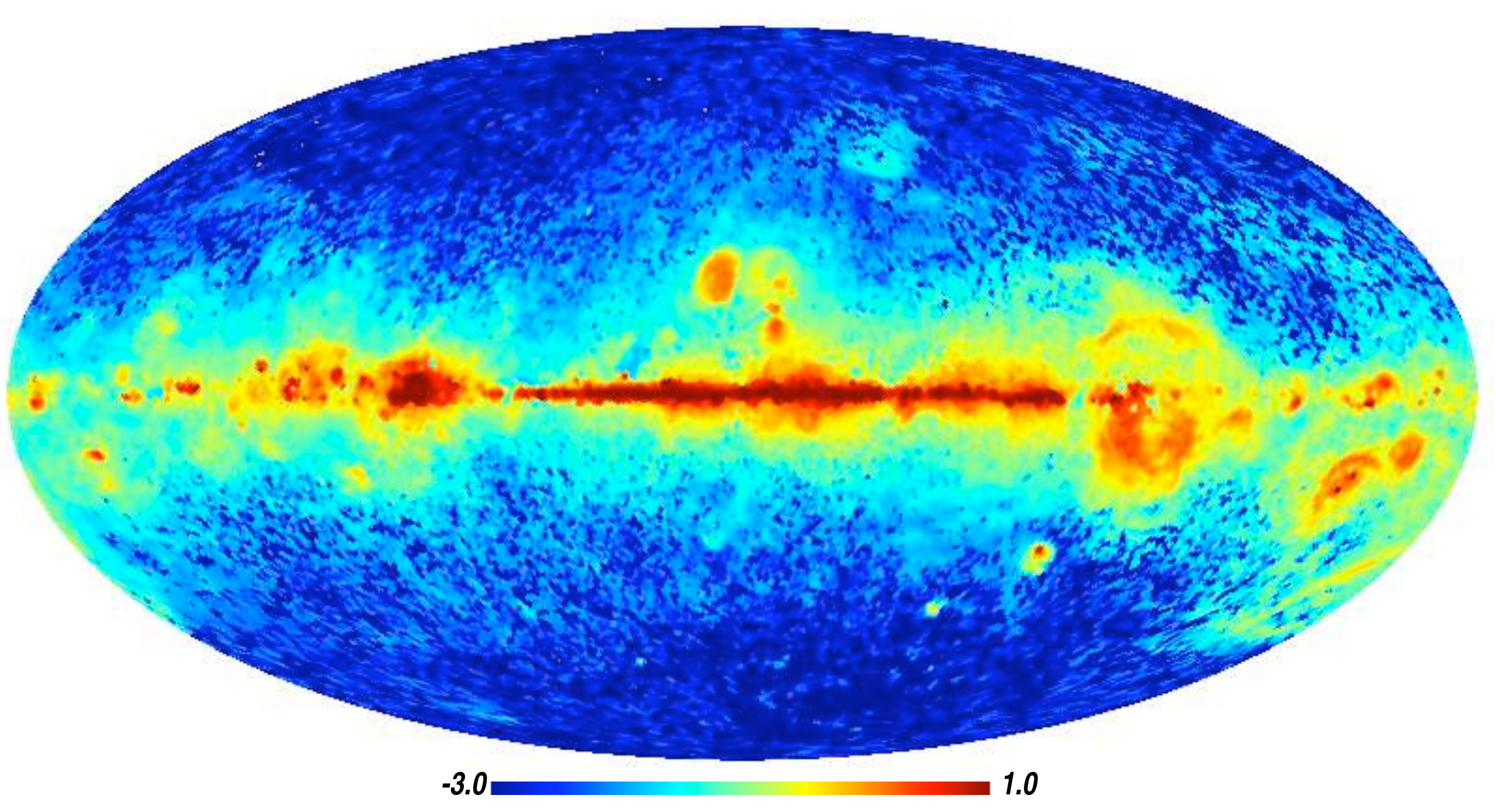}
\caption{\label{fig:freefree} Free-Free emission at 23 GHz (in log$_{10}$(mK)) 
estimated using H$\alpha$ emission
corrected for extinction (following \cite{dickinson2003}) and the WMAP MEM decomposition.
For $E(B-V) \ge 2$, the free-free emission is the one obtained from the WMAP MEM decomposition.
For $E(B-V) < 2$, the free-free emission is the one estimated using H$\alpha$ emission, expect
when the WMAP MEM free-free is lower.
}
\end{figure}

In this paper we want to put some constraints on the presence and properties of the 
anomalous emission in the WMAP frequency range,
while making minimum assumptions about the emission mechanisms. 
The difficulty of this task comes from our lack of knowledge on the other components at these frequencies, 
mostly synchrotron.
Anomalous emission has been measured to be very weakly polarized by \cite{Battistelli2006}.
This can be understood if the carriers of the spinning dust emission are small dust grains 
which are known not to produce any polarization from extinction measurements \cite[e.g.][]{martin2007}.
Free-free is also known to produce no polarization\footnote{expect at the edge of
HII regions}. We therefore follow \cite{finkbeiner2004} and use polarization data
to estimate the synchrotron emission. 
Based on this assumption and on a model of the Galactic magnetic field, 
we use the WMAP 23~GHz polarization data to put realistic 
constraints on the synchrotron intensity and isolate the anomalous emission component.

The paper is organized as follow: the WMAP data we used are summarized in \S~\ref{sec:data}.
Then \S~\ref{sec:models} describes the models we used for the Galactic components,
\S~\ref{sec:polarization} describes our model for the synchrotron polarized data
and for the Galactic magnetic field, and \S~\ref{sec:anomalous} presents 
the anomalous emission extracted from the WMAP data. The results are discussed and summarized
in \S~\ref{sec:conclusion}.

\section{WMAP data}

\label{sec:data}

In this analysis we use the three-year WMAP products available in the Healpix pixelisation scheme
\cite[]{gorski2005} on the LAMBDA website\footnote{http://lambda.gsfc.nasa.org}.
We use the temperature and polarization (I, Q, U) coadded maps
per frequency in the K (23~GHz), Ka (33~GHz), Q (41~GHz), V (61~GHz) and W (94~GHz) bands.
The original angular resolution of each map ranges from $0.23^\circ$ at 94~GHz
to $0.93^\circ$ at 23~GHz. The maps were all smoothed to a common resolution of 1$^\circ$.
We also use the Galactic Kp2 and point sources masks provided by the WMAP team.

To remove the Cosmic Microwave Background (CMB) from the data we use the estimate based
on the Internal Linear Combination (ILC) method proposed by the WMAP team and
applied on the three years data \cite[]{bennett2003}. The resolution of the ILC map is also $1^\circ$.
As shown by \cite{davies2006,dobler2007} the estimate of the CMB can have a significant impact on the
determination of the foreground spectra. In our case, it is a sub-dominant effect as 
our analysis is conducted at 23~GHz where the Galaxy-to-CMB ratio is the highest.

By design WMAP is unable to measure the zero level at each frequency. This has no consequence
for the study of the CMB fluctuations but it has an importance for the determination of the foreground properties.
In practice there is a degeneracy between the zero level and 
the absolute foreground component amplitudes one can deduced from the data, unless one relies
on a correlation analysis like \cite{davies2006} who measured a mean spectral index of 
$\beta_s\approx -3$ between 408~MHz and 23~GHz at high Galactic latitude. 
In their analysis \cite{eriksen2007} fitted jointly the CMB, foregrounds and zero level.
In order to estimate the zero level they put a constrain on the average value of $\beta_s$ based on the results of \cite{davies2006}.
In our analysis we rely on the zero level determined by \cite{eriksen2007}. 
In practice this zero level correction is only relevant at 23~GHz.

\begin{figure}
\includegraphics[width=\linewidth, draft=false]{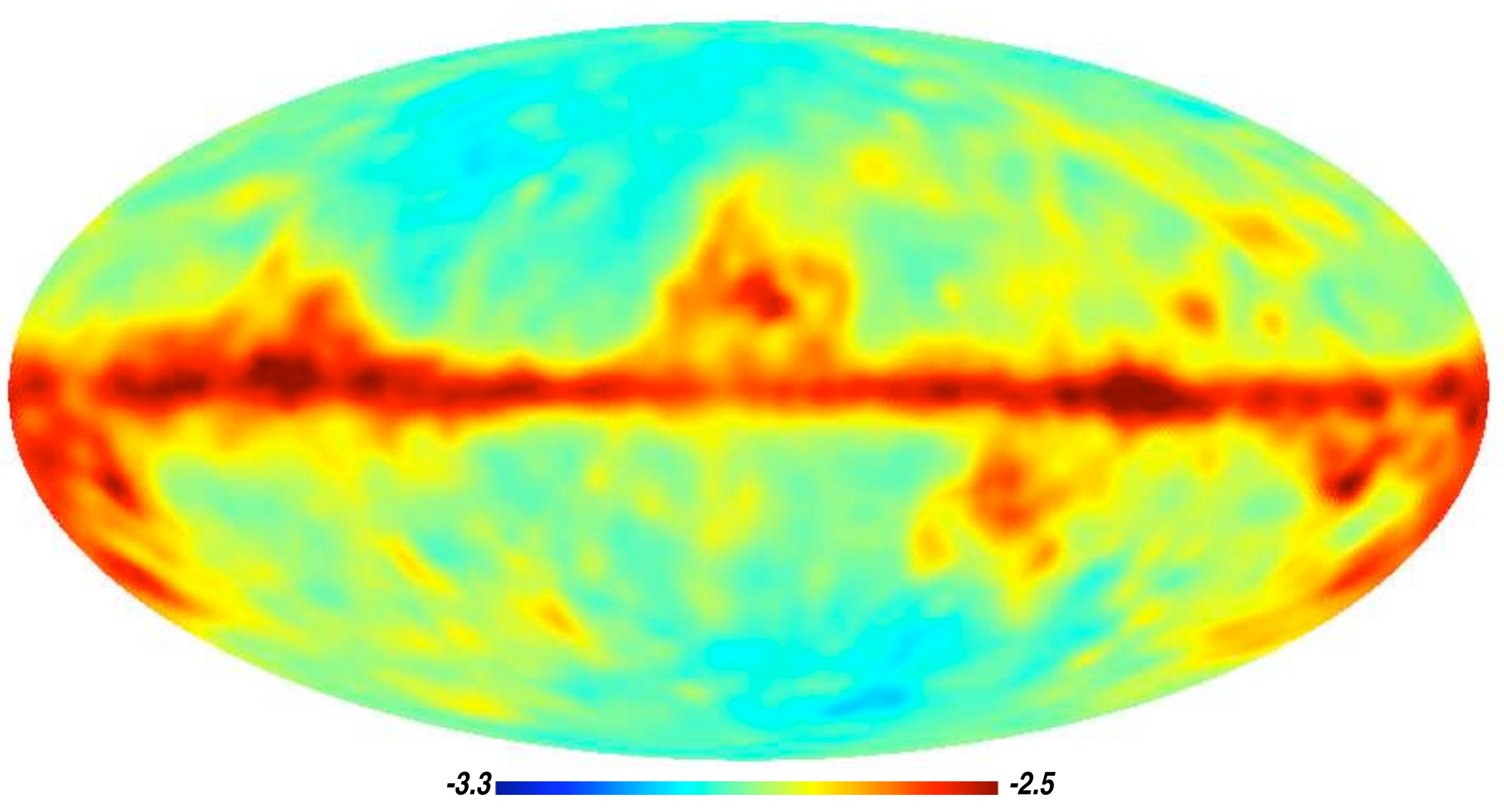}
\includegraphics[width=\linewidth, draft=false]{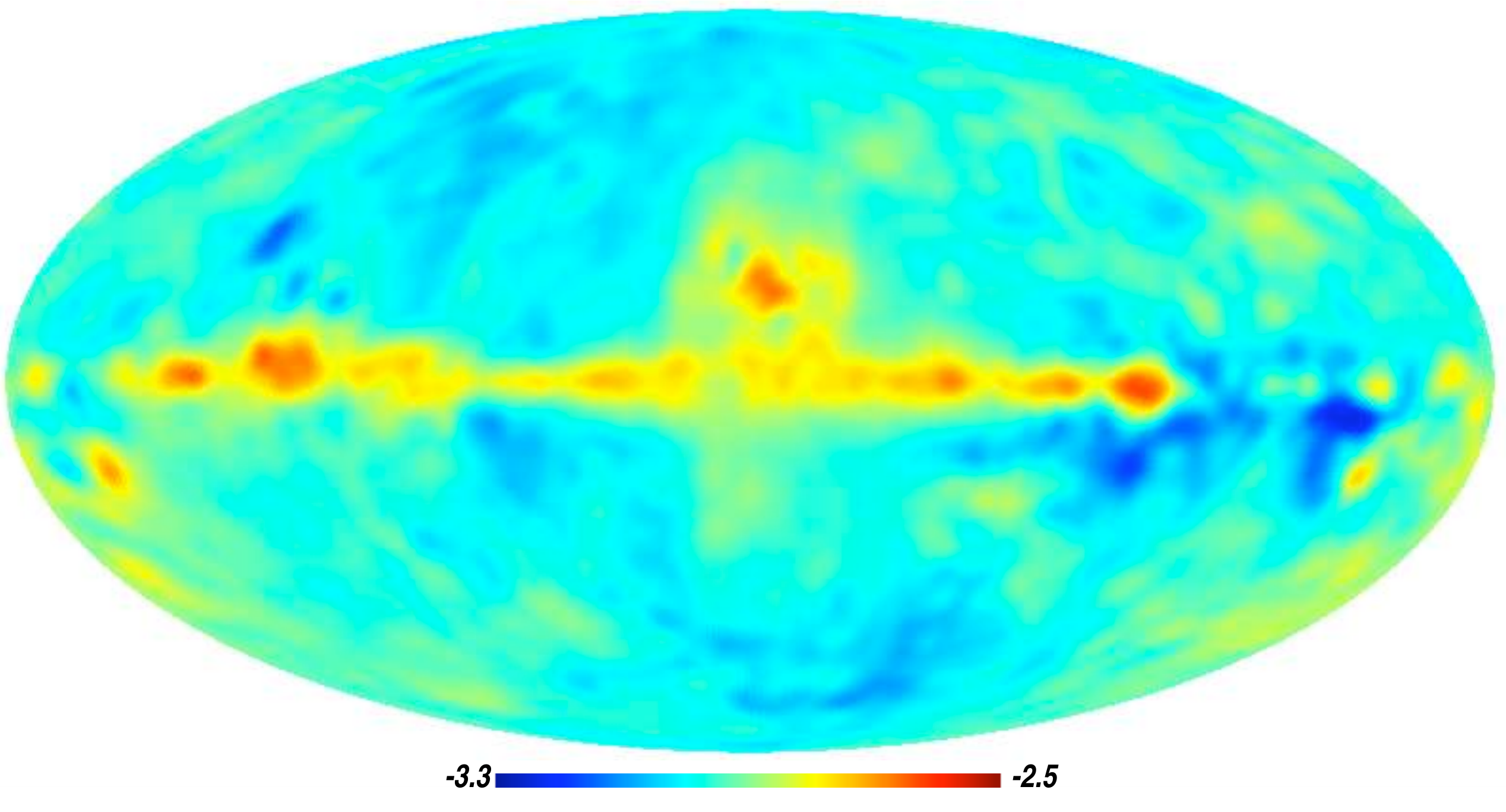}
\includegraphics[width=\linewidth, draft=false]{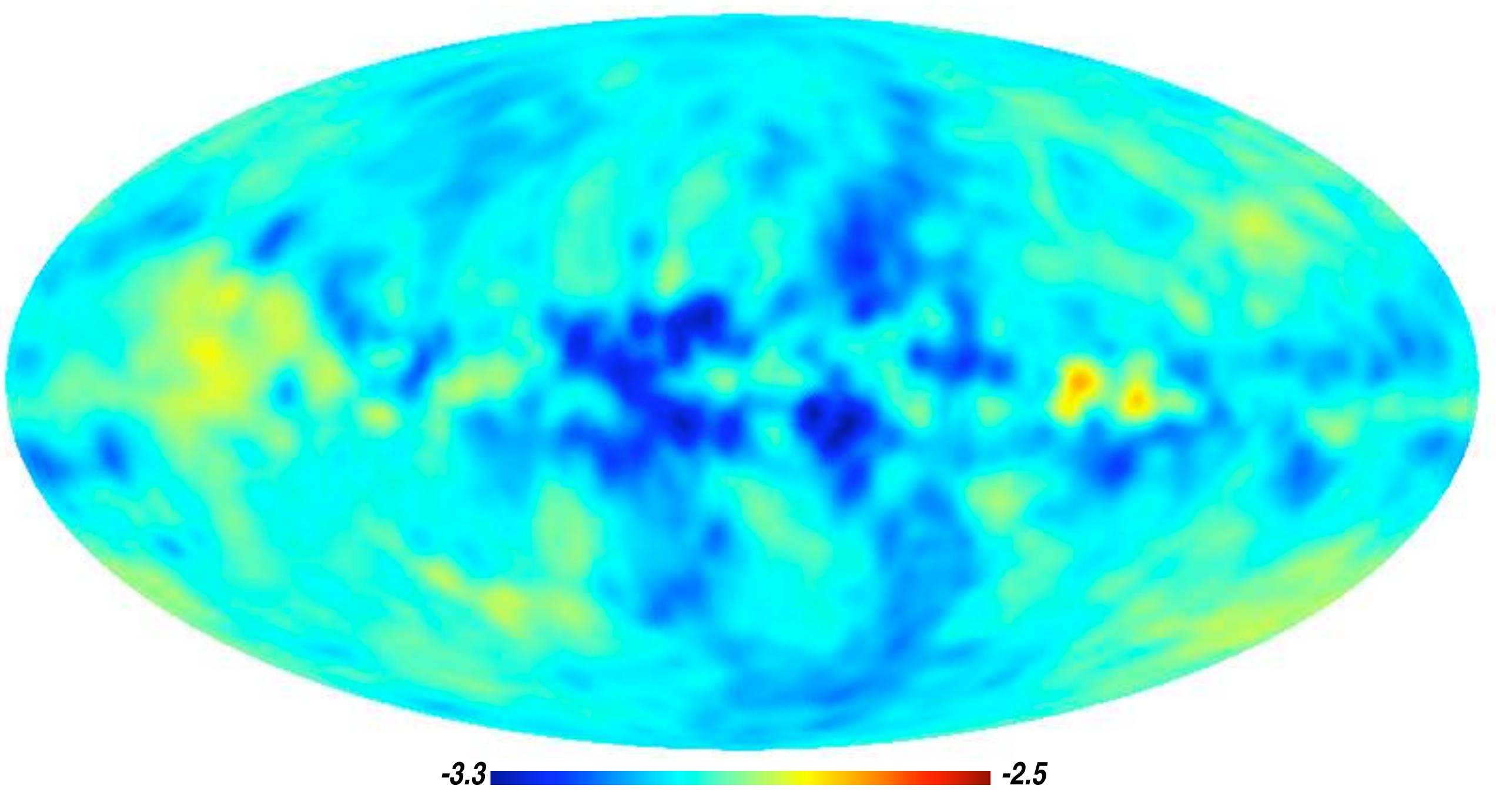}
\caption{\label{fig:beta} Synchrotron spectral index ($\beta_s$) between 408~MHz and 23~GHz.
All maps were smoothed at $5^\circ$.
{\bf Top:} Model 1 - this is making the assumption of no anomalous emission.
The map is similar to the one of \cite{bennett2003a} (their Fig.~4). 
{\bf Middle:} Model 3 - assumes a spinning dust component at 23 GHz which scales with E(B-V). 
{\bf Bottom:} Model 4 - $\beta_s$ obtained from polarization data for 
$p=-8.5^\circ$, $\chi_0=8^\circ$, $r_0=11^\circ$, and 
$\sigma_{turb}=1.7$~$\mu$G.}
\end{figure}

\begin{figure}
\includegraphics[width=\linewidth, draft=false]{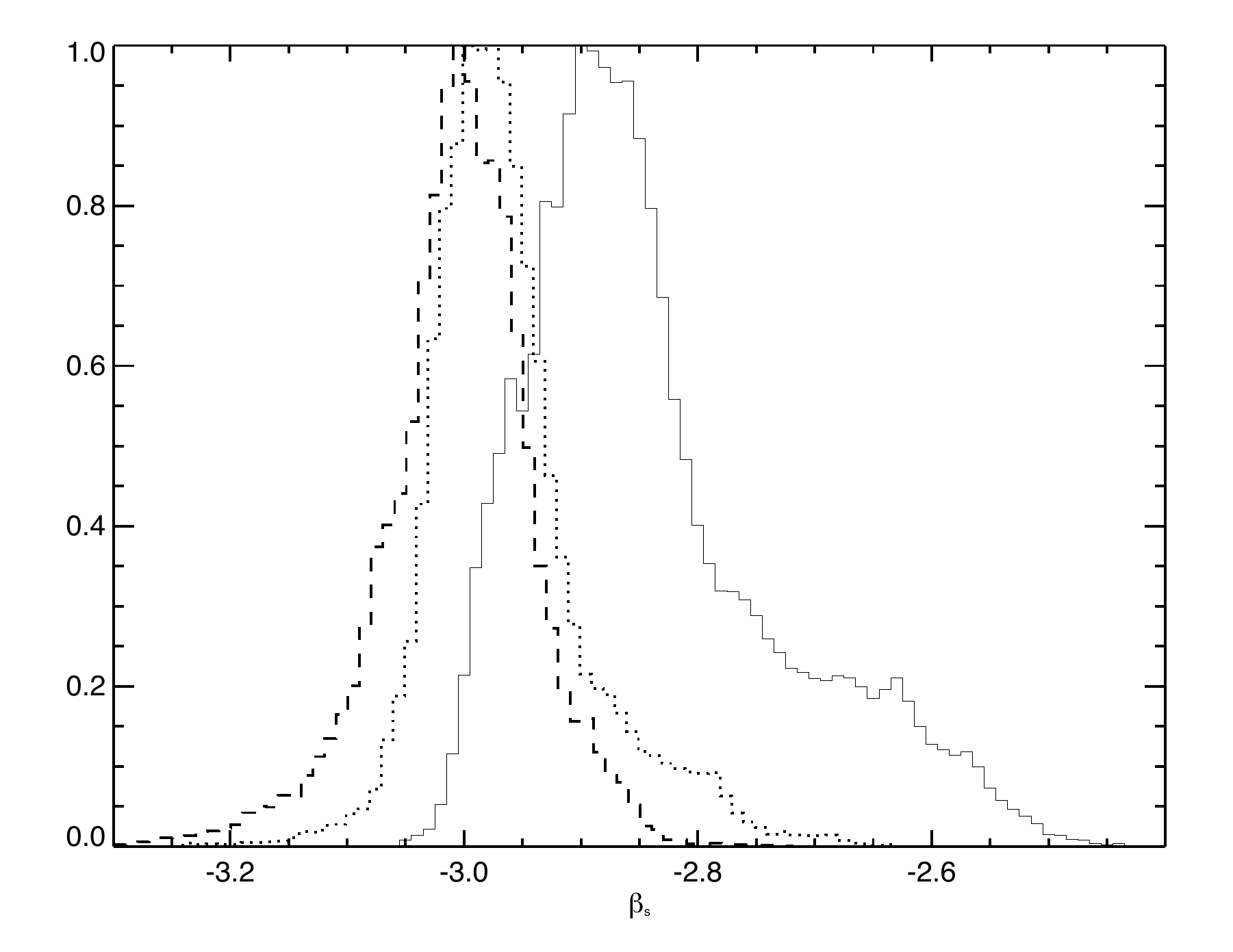}
\caption{\label{fig:histo_beta} Histogram of $\beta_s$ between
 408 MHz and 23 GHz smoothed at 5$^\circ$. {\bf solid line:} model 1, $\beta_s=-2.83\pm0.11$ 
 {\bf dotted line:} model 3, $\beta_s=-2.96\pm0.07$ 
 {\bf dashed line:} model 4,
 $\beta_s = -3.00\pm0.06$.}
\end{figure}

\begin{figure}
\includegraphics[width=\linewidth, draft=false]{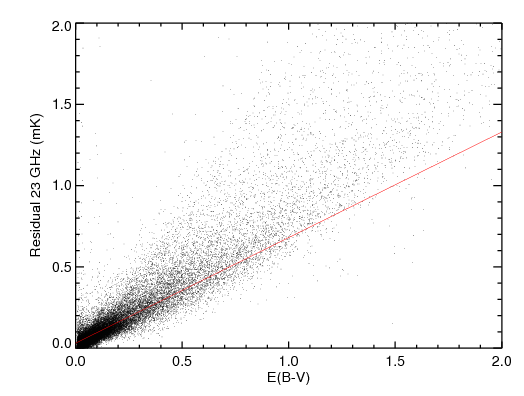}
\caption{\label{fig:correl_ebmv_residu} Residual 23 GHz emission when free-free and
synchrotron with a constant $\beta_s=-3$ are subtracted, as a function of E(B-V).
The solid line is 0.65 E(B-V) + 0.03.}
\end{figure}

\section{Galactic components}

\label{sec:models}

\subsection{WMAP data model}

In the following we consider that 
the WMAP intensity $I_\nu$ (corrected for CMB, zero level and dipole, and smoothed at 1$^\circ$) 
at frequency $\nu$ is the sum of free-free ($F_\nu$), 
synchrotron ($S_\nu$), thermal dust ($D_\nu$) and anomalous emissions ($A_\nu$)\footnote{in one model (model \#1) we
assume no anomalous emission}:
\begin{equation}
I_\nu = S_\nu + F_\nu + D_\nu + [A_\nu].
\end{equation}

In polarization, the model of the WMAP data involves less components
as the free-free and anomalous emissions are assumed to be unpolarized. 
The CMB polarization level is so weak \cite[]{page2007} that it is not included in the data model.
The polarized emission model is thus simply:
\begin{equation}
P_{\nu} = S_{\nu}^P + D_\nu^P.
\end{equation}

\subsection{Free-Free}

\label{section:free-free}

Free-free emission is a rather difficult component of the interstellar emission
to identify as it does not dominate at any frequency. H$\alpha$, corrected
for dust extinction, is often used as a proxy for free-free as they both depend linearly on the 
emission measure, EM=$\int n_e^2 \, dl$ (see \cite{dickinson2003} and \cite{finkbeiner2003}).
On the other hand, several systematic errors can affect the estimate of free-free emission from $H\alpha$ corrected
for extinction (see \cite{bennett2003} and \cite{finkbeiner2004}).
\cite{finkbeiner2004} estimated that an error of a factor of 2 
is likely in some parts of the sky solely due to an imperfect extinction correction. 
Electronic temperature variations are also expected, with lower values in the Galactic plane due to a higher metallicity
and therefore more efficient cooling. Estimate of spatial variations of $T_e$ are 
not available and a constant value on the sky is usually assumed.

It is important to point out that the free-free spectral index is a 
slowly varying function of frequency and electron temperature
\cite[]{bennett1992,dickinson2003,bennett2003a}:
\begin{equation}
\beta_{ff} = 2+ \frac{1}{10.48 + 1.5 \ln(T_e/8000K) - \ln(\nu_{GHz})}.
\end{equation}
At 23~GHz, the free-free spectral index varies only from 2.13 to 2.16 for $4000 < T_e < 10000$~K
and it is not affected by extinction.
Based on the analysis by \cite{davies2006} we also note that, at this frequency, 
the spectral index of free-free is significantly different than
that of synchrotron ($\beta\sim-3$) and anomalous emission ($\beta\sim-2.85$).
This is a favorable configuration for a maximum entropy method (MEM) analysis
that can easily isolate a component over the whole sky that has 
a fixed spectral index using five frequency bands, especially
if the other components have significantly different spectral indices. 

Based on this physical property of the free-free emission \cite{bennett2003a,hinshaw2007} estimated the
free-free emission in the WMAP data using MEM and considering a constant spectral index of 2.14
over the whole sky.
In the following, we rely on the result of the MEM decomposition of \citet{bennett2003a}
for the estimate of the free-free emission in regions where $E(B-V) > 2$ (or $A_V \geq 6$).
The same assumption guided the work of \cite{sun2007a}.
In more diffuse regions we use the free-free estimate of \cite{dickinson2003} based on 
a H$\alpha$ all-sky template at 1$^\circ$ resolution, unless the WMAP MEM free-free is lower than
the H$\alpha$ estimate. The free-free template we use at 23~GHz is shown in Fig.~\ref{fig:freefree}.

\subsection{Thermal dust}

To estimate thermal dust emission in the WMAP data we used model \#7 of \cite{finkbeiner1999}, multiplied
by 1.2 to account for the emission in the 94~GHz band \cite[]{hinshaw2007}.
The correlation of the predicted thermal dust emission with $(I_{94}-F_{94})$ at 94~GHz 
and inside the Kp2 mask is 0.96.
We note that this component contributes only to $\sim 1\%$ of the emission at 23~GHz
and that it does not affect our analysis of the synchrotron and anomalous emission at this frequency.
This is true also in polarization as the polarization fraction of thermal dust is 
at the level of only a few percent \cite[]{benoit2004,ponthieu2005,page2007}. 
Therefore at 23~GHz, where most of our analysis is done, 
we considered the polarized intensity to be entirely due to synchrotron emission: $P_{23}=S_{23}^P$.
 
\subsection{Synchrotron}

\label{section:synchrotron}

The synchrotron intensity depends on the cosmic ray (electron) density ($n_e$)
and the strength of the magnetic field perpendicular to the line of sight ($B_\perp$).
For a cosmic ray distribution following a power law, $N(E) \propto E^{-s}$, 
the synchrotron intensity at frequency $\nu$ is:
\begin{equation}
\label{eq:I_sync}
S(\nu) = \epsilon_s(\nu) \int_z n_e B_\perp^{(1+s)/2} \, dz
\end{equation}
where the integral is over the line of sight $z$ and $B_\perp = \sqrt{B_x^2 + B_y^2}$
with the plane $x-y$ corresponding to the plane of the sky.
The emissivity term $\epsilon_s(\nu)$ is given by a power law:
\begin{equation}
\epsilon_s(\nu) = \epsilon_0 \nu^{\beta_s}
\end{equation}
where 
\begin{equation}
\label{eq:betas}
\beta_s =-(s+3)/2.
\end{equation}
It follows that the synchrotron intensity spectrum is a power law
\begin{equation}
S(\nu) = S(\nu_0) \left(\frac{\nu}{\nu_o}\right)^{\beta_s}.
\end{equation}
For a typical value of cosmic ray spectrum $s=3$, we obtain $\beta_s=-3$.

The most reliable estimate of synchrotron emission to date is the 408~MHz all-sky map
of \cite{haslam1982}. 
In the following we estimate the synchrotron emission at any WMAP frequency 
using an extrapolation of the Haslam 408~MHz map ($S_{408}$) 
corrected for free-free emission as provided by the WMAP team:
\begin{equation}
S_{\nu} = S_{408}\left(\frac{\nu}{0.408}\right)^{\beta_s}.
\end{equation}

The extrapolation of the 408~MHz emission at WMAP frequencies relies on an estimate of $\beta_s$ and
of its variations on the sky.
All-sky estimates of the synchrotron spectral index between 408 MHz and 2326 MHz
have been attempted by \cite{giardino2002,platania2003}. They obtained compatible
values of $\beta_s = -2.7 \pm 0.1$. 
On the other hand the spectral index varies with frequency \cite[]{banday1991}
due to electron aging effects. This effect is expected to be 
important in the WMAP range which 
makes uncertain the extrapolation to 23~GHz of the spectral index defined 
at lower frequencies. 

From the analysis of the WMAP data \cite{finkbeiner2004} found that a constant $\beta_s=-3.05$ 
between 408~MHz and 23~GHz corrects the most obvious synchrotron features seen both at 408~MHz and 23~GHz (i.e. Loop 1).
This is in accordance with the work of \cite{davies2006} who obtained $\beta_s=-3.01 \pm 0.04$ outside the Kp2 mask.

In the following we describe four different models of the synchrotron spectral index
between 408~MHz and 23~GHz.
In all models we considered that thermal dust emission 
is negligeable at 23~GHz ($D_{23}\sim0$) and that the free-free ($F_\nu$)
is given by our model (see \S~\ref{section:free-free}).

\subsubsection{Model 1}

Following \cite{bennett2003}, this model assumes 
that the 23~GHz emission is only synchrotron and free-free, giving
an upper limit on the synchrotron spectral index:
\begin{equation}
\label{eq:beta_max}
\beta_s = 0.248\, \log\left(\frac{I_{23}-F_{23}}{S_{408}}\right)
\end{equation}
where the multiplicative factor 0.248 is simply $1/\log(23/0.408)$.
The map of $\beta_s$ obtained is shown in the top panel of Fig.~\ref{fig:beta}.
Using this simple method, which does not take into account anomalous emission, 
we obtain $\beta_s= - 2.84\pm0.12$. The histogram of $\beta_s$ is shown in 
Fig.~\ref{fig:histo_beta} (solid line).
Taking into account anomalous emission will lower the contribution of synchrotron at 23 GHz
which will increase the value of $\beta_s$.
This $\beta_s$ is similar to the one obtained by \cite{bennett2003}. The differences
are mainly due to the zero level correction we applied and to our slightly different free-free estimate.

\subsubsection{Model 2}

In this model we assume the presence of anomalous emission making no hypothesis on its
spatial distribution and we set a constrain on $\beta_s$ considering it constant 
with a value of $\beta_s=-3$ in accordance with \cite{davies2006}:
\begin{equation}
I_{23} = S_{408}\left(\frac{23}{0.408}\right)^{-3} + F_{23} + A_{23}.
\end{equation}
When this synchrotron estimate and our free-free model are subtracted, the residual ($A_{23}$)
is shown to be well correlated to the estimate of E(B-V) of \cite{schlegel1998} smoothed at 1 degree.
In diffuse regions (E(B-V)$<0.5$) we obtained $A_{23} \sim (0.65\pm0.01)$~E(B-V)~$+(0.03\pm0.01)$ 
(see Fig.~\ref{fig:correl_ebmv_residu}). 

Even though the estimate of E(B-V) of \cite{schlegel1998} has been shown to overestimate systematically
the extinction in regions where colder dust is present \cite[see ][]{cambresy2001,cambresy2005}, 
it remains a relatively accurate tracer of dust column density in diffuse regions, typically where E(B-V)$<0.5$.

\subsubsection{Model 3}

We can improve the estimate of $\beta_s$ and estimate its spatial variation 
by assuming a template for the anomalous emission component. Here we make the assumption that
the anomalous emission is well correlated with dust column density (or E(B-V)). We added a component
that scales with E(B-V): $I_{23} = S_{23} + F_{23} + a\, E(B-V)$.
Here the synchrotron spectral index is:
\begin{equation}
\beta_s = 0.248 \, \log\left( \frac{I_{23}-F_{23}-a\,E(B-V)}{S_{408}}\right).
\end{equation}

In this simple model we use $a=0.65$~mK/mag which is the slope observed in diffuse regions in model~2
(see Fig.~\ref{fig:correl_ebmv_residu}).
Compare to model~1, this model (see Fig.~\ref{fig:beta}, middle panel and 
Fig.~\ref{fig:histo_beta} (dotted line)) produces 
a steeper $\beta_s$ on average ($\beta_s = -2.96 \pm 0.07$) but compatible with model~2 (constant $\beta_s=-3$)
which is expected as we used the value of $a$ deduced from model~2.
Most importantly the spatial variations of $\beta_s$ is reduced by a factor of 2 compared to model~1.
This has to be appreciated considering 
that some spatial variations of $\beta_s$ could be due to spatial variations of the
anomalous emission that we considered to be strictly proportional to $E(B-V)$.
Nevertheless the fact that taking into account an extra component reduces significantly the variations of $\beta_s$ 
is a strong indication of a better modeling of the data.
This is compatible with gamma ray observations 
from which \cite{strong2000} do not predict strong spatial variations of the synchrotron spectral index,
mostly due to the large interaction length of cosmic ray electrons.

\subsubsection{Model 4}

In order to separate the synchrotron and anomalous emissions at 23 GHz we would like
to have an estimate of $\beta_s$ independent of intensity measurements at 23 GHz
and of any constrains on the spatial variation of the anomalous emission.
In model 4 we use the fact that the polarized intensity at 23~GHz is dominated 
by synchrotron. Based on a model of the Galactic magnetic field we estimated
the synchrotron intensity at 23~GHz using the WMAP polarization data.
This model is described in the next section.

\section{Synchrotron polarization at 23 GHz}

\label{sec:polarization}

\subsection{Synchrotron polarization}

Similarly to the synchrotron intensity (see Eq.~\ref{eq:I_sync}) 
the stokes parameters $Q$ and $U$ of polarized synchrotron emission
integrated along the line of sight are 
\begin{equation}
\label{eq:Q_sync}
Q(\nu) = f_s \epsilon_s(\nu) \int_z n_e B_\perp^{(1+s)/2} \cos 2\phi \sin \alpha \, dz
\end{equation}
and
\begin{equation}
\label{eq:U_sync}
U(\nu) = f_s \epsilon_s(\nu) \int_z n_e B_\perp^{(1+s)/2} \sin 2\phi \sin \alpha \, dz
\end{equation}
where
\begin{equation}
\cos 2 \phi = \frac{B_x^2-B_y^2}{B_\perp^2},
\end{equation}
\begin{equation}
\sin 2 \phi = \frac{-2B_xB_y}{B_\perp^2},
\end{equation}
and
\begin{equation}
\sin \alpha = \sqrt{1-B_z^2/B^2}.
\end{equation}
The polarization fraction $f_s$ is related to the cosmic ray energy distribution slope $s$:
\begin{equation}
\label{eq:fs}
f_s = \frac{s+1}{s+7/3}.
\end{equation}
For a typical value of cosmic ray spectrum $s=3$ we obtain 
$f_s=0.75$.

The polarized intensity is simply
\begin{equation}
\label{eq:P_sync}
P(\nu) = \sqrt{Q(\nu)^2 + U(\nu)^2}
\end{equation}
and the polarization angle
\begin{equation}
\gamma(\nu) = 0.5\tan^{-1}\left( \frac{U(\nu)}{Q(\nu)}\right).
\end{equation}

\begin{figure}
\includegraphics[width=\linewidth, draft=false]{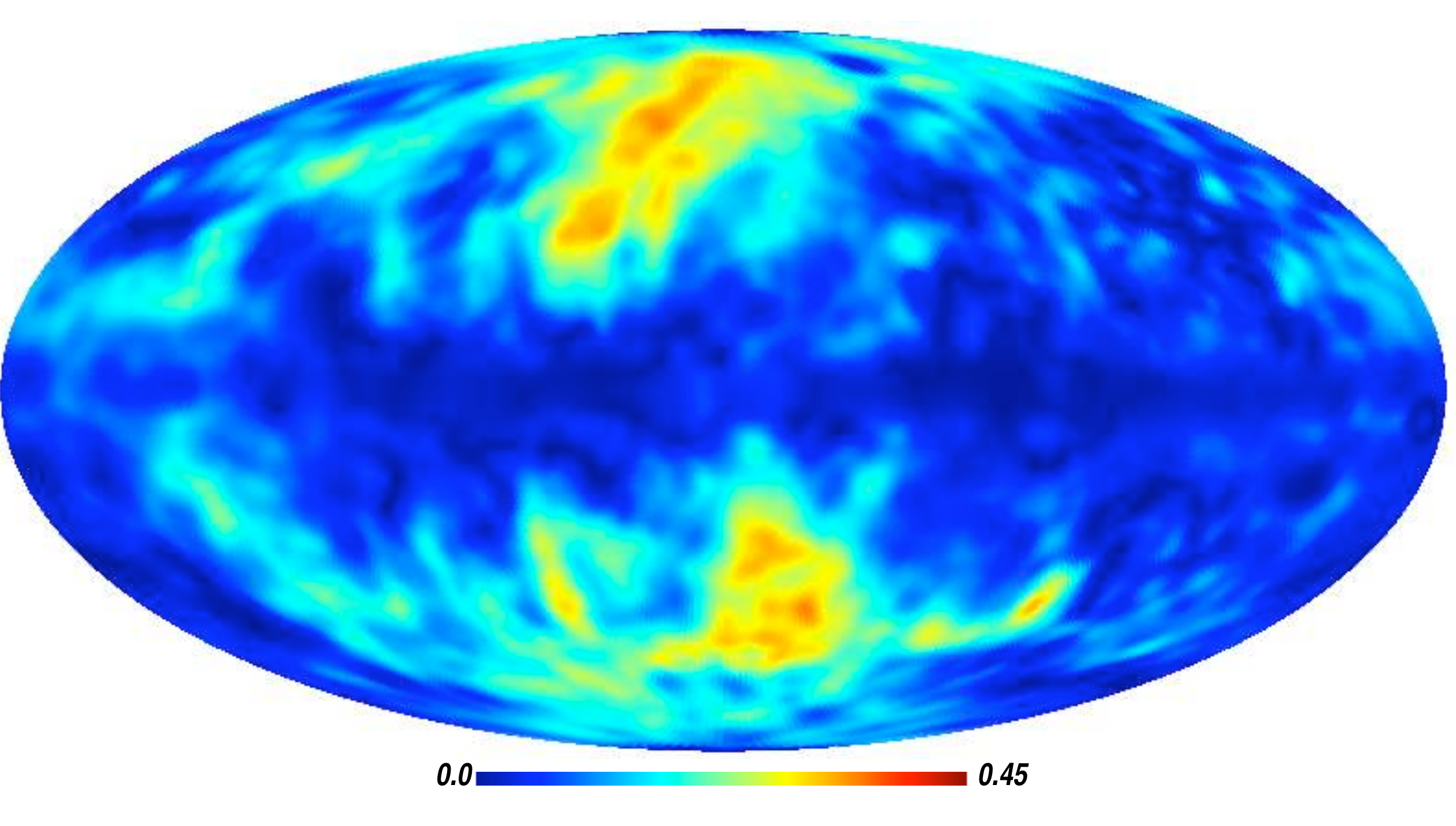}
\includegraphics[width=\linewidth, draft=false]{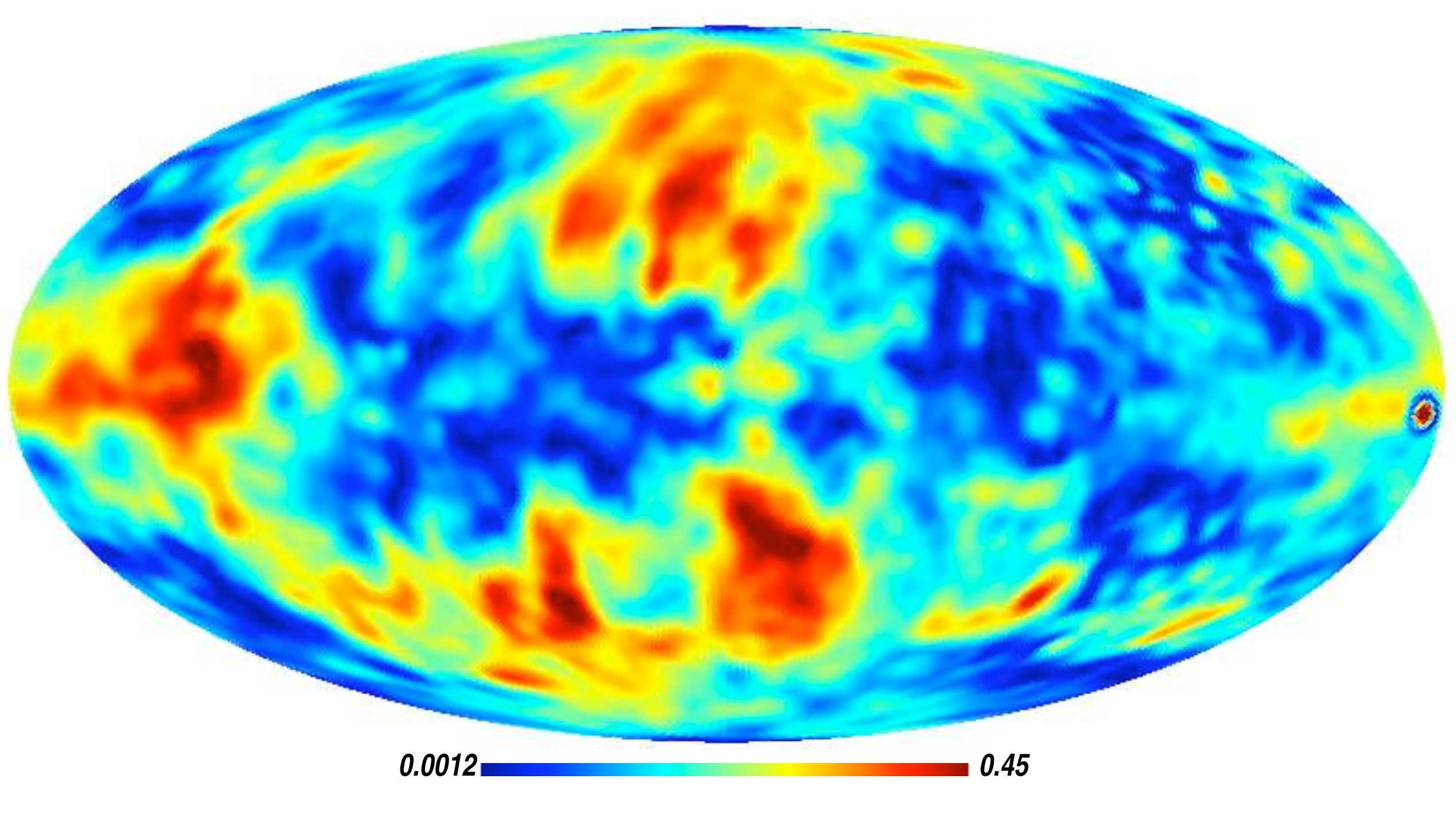}
\includegraphics[width=\linewidth, draft=false]{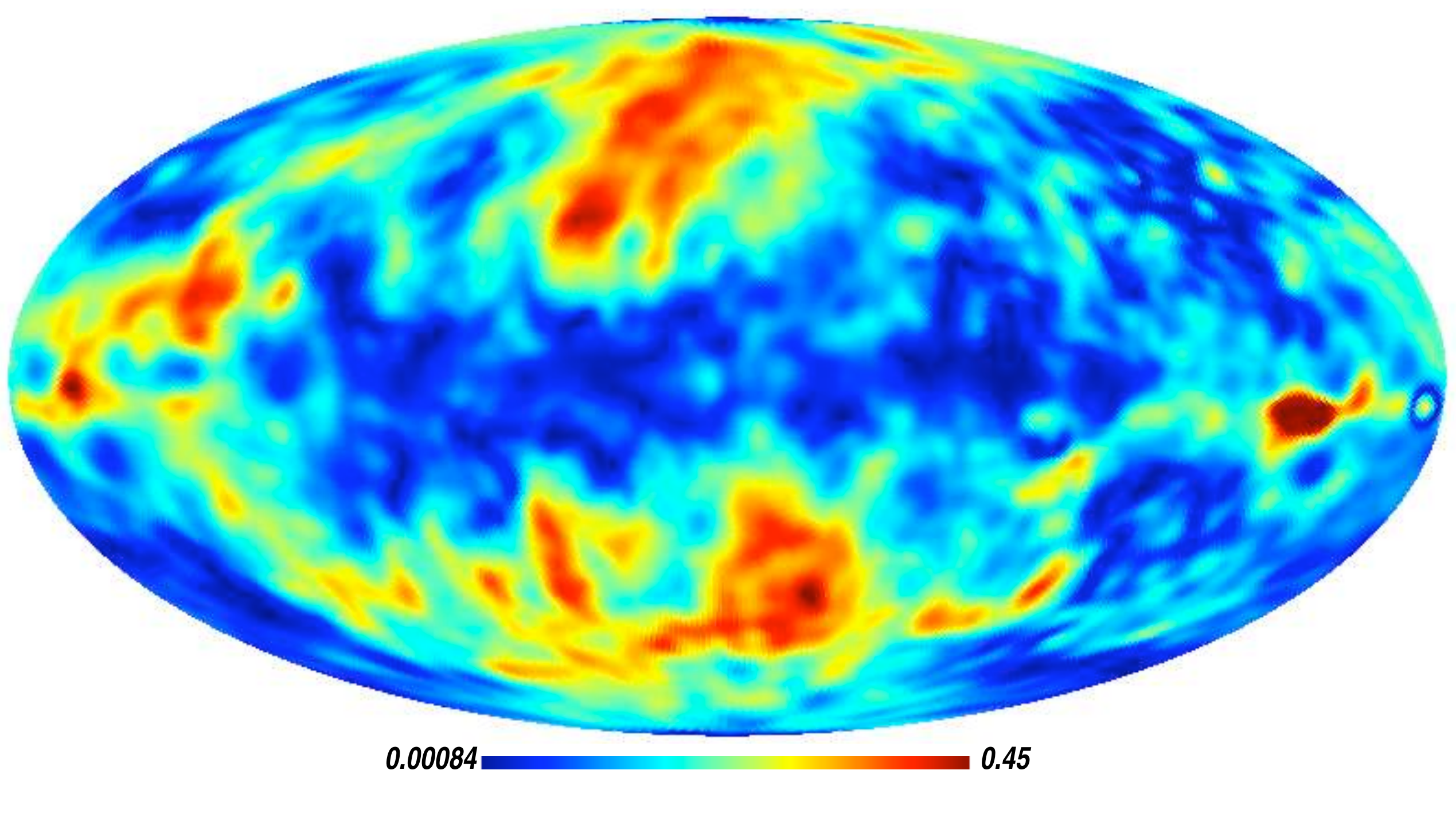}
\caption{\label{fig:polar_fraction} Polarization fraction at $5^\circ$ resolution.
The polarization $P$ is from the WMAP K band data and
the intensity $S$ is from models 1 to 3 (from top to bottom).}
\end{figure}

\subsection{Polarization fraction at  23 GHz}

Combining equations~\ref{eq:I_sync}, \ref{eq:Q_sync} and \ref{eq:U_sync},
the polarization fraction is given by the following simple equation: 
\begin{equation}
\frac{P(\nu)}{S(\nu)} =  f_s \, g
\end{equation}
where $f_s$ is the intrinsic polarization fraction related to the cosmic rays energy spectrum (see Eq.~\ref{eq:fs})
and $g$ is a geometrical reduction factor that reflects the depolarization due to the variation of the angle between 
the magnetic field and the line of sight directions.
The $g$ factor depends on the structure of
the magnetic field and on the distribution of the cosmic rays in three-dimensions
(i.e. the integrals in equations~\ref{eq:I_sync}, \ref{eq:Q_sync} and \ref{eq:U_sync}).
It takes values between 0 and 1; it reaches a maximum when the magnetic field is
parallel to the plane of the sky ($B=B_\perp$). In practice it never reaches 1 as the
magnetic field direction fluctuates on any line of sight, due to both the
spiral structure of the magnetic field and turbulent motions.

We can estimate the polarization fraction at 23~GHz using the WMAP polarization data ($P_{23}$)
and estimates of the synchrotron intensity ($S_{23}$) given by models 1 to 3. 
The corresponding polarization fraction maps, smoothed at $5^\circ$ 
are shown in Fig.~\ref{fig:polar_fraction}.
The polarization fraction for model 1 is similar to the one 
given by \cite{kogut2007} (their Fig.~4). 
Models 2 and 3 gives very similar polarization fraction maps, which is expected 
considering that both models have
almost the same average $\beta_s$ and considering the small dispersion of $\beta_s$ in model~3.
On the other hand the polarization fraction of model 1 is systematically lower than in other models because of the assumption of
no anomalous emission which increases the synchrotron intensity (and not its polarization). 
Apart from the average level, the most significant difference between the 
polarization fraction of model 1 to 3 is the large scale structure. 
Models 2 and 3 shows less polarization in the inner galaxy than towards the anti-center region. 
Model 1 has a systematically lower polarization value in the plane over all longitudes
without a clear raise of the polarization fraction towards the anti-center. 
The contrast between low and mid latitudes is also stronger in model 1 than in models 2 and 3.

The structures seen in the polarization fraction maps are due to real depolarization structure
(due to the magnetic field 3D structure - large scale and turbulent components)
but also to our imperfect  estimate of $\beta_s$. Nevertheless it is remarkable that
the polarization fraction given by models 2 and 3 are very similar even with important
differences in the definition of $\beta_s$. In the following 
we look for a magnetic field model that allows to reproduce these polarization fraction maps.

\subsection{Model of polarization at 23 GHz}

In model~4 we use the relation between intensity and polarization to
estimate the synchrotron intensity from the 23~GHz polarization data which are
assumed to be dominated by synchrotron polarization: 
\begin{equation}
S_{23} \sim \frac{P_{23}}{g f_s}.
\end{equation}
The synchrotron spectral index is then given by:
\begin{equation}
\beta_s = 0.248 \, \log\left(\frac{P_{23}}{g \, f_s\, S_{408}}\right).
\end{equation}
This relies on a proper estimate of the reduction factor $g$.
The construction of $g$ depends on a model of the large scale galactic magnetic field
and of its fluctuations due to turbulence. The latter will
have a strong impact on the depolarization value, especially in the Galactic plane
where lines of sight are longer.

In order to estimate $g$, we use the fact that $g \sim P/S$ ($f_s$ is
almost constant - see Eq.~\ref{eq:fs}). To estimate the polarization fraction $P/S$
we use equations \ref{eq:I_sync}, \ref{eq:Q_sync}, \ref{eq:U_sync} and \ref{eq:P_sync} which depends
on the structure of the magnetic field $B$ in three dimensions and on the electronic density $n_e$.
Following \cite{page2007} we used a variation of the cosmic ray electron density with galacto-centric
radius and height:
\begin{equation}
n_e = n_0 \exp(-r/h_r) \, \mbox{sech}^2(z/h_z)
\end{equation}
with $h_r=5$~kpc and $h_z=1$~kpc. 
\cite{sun2007a} adopted a similar description of the cosmic ray distribution.

\subsubsection{Galactic magnetic field at large scales}

Following what is found in the analysis of pulsar rotation measures (RM) \cite[]{han2006b} 
we modeled the Galactic magnetic as a bi-symmetrical spiral (BSS):
\begin{equation}
B(r,\theta, z) = B_0(r) \cos(\theta - \psi\ln\frac{r}{r_0} ) \cos(\chi)
\end{equation}
where $\psi = 1/\tan(p)$, $p$ being the pitch angle
and $\chi(z) = \chi_0 \tanh(z/\mbox{1 kpc})$
describing the vertical ($z$) component of the magnetic field \citep[see][]{page2007}.
Using RMs at high Galactic latitudes, \cite{han2006b} estimated
that the amplitude of this component of the magnetic field is almost
10 times smaller than the horizontal components. 
We found that $\chi_0$ has little impact on the modeling.

Rotation measures are compatible with a slight variation of the magnetic field strength $B_0$
with Galactic radius $R$, modeled by \cite{han2006b} with a slowly rising exponential 
towards the Galactic center. 
In fact this increase is not very clear and the data are also compatible 
with $B_0$ constant with $R$. In this analysis we found that a constant 
$B_0$ provides a better fit to the WMAP data.

\begin{figure}
\includegraphics[width=\linewidth, draft=false]{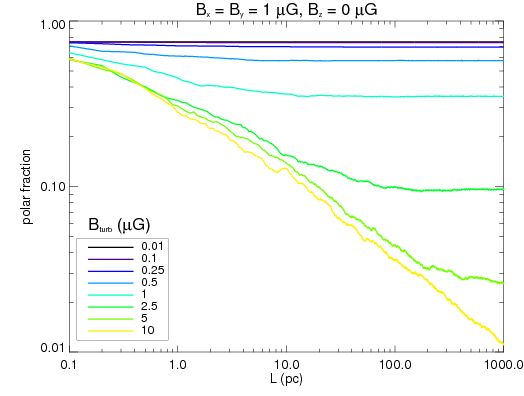}
\caption{\label{fig:turbul} Polarization fraction as a function of 
depth along the line of sight. The large scale magnetic field is perpendicular
to the line of sight, giving maximum polarization ($g=1$ and $f_s=0.75$) with no turbulence.
The different curves give the level of polarization we would observe for increasing
level of the turbulent component of the magnetic field. }
\end{figure}

\subsubsection{Turbulent part of the magnetic field}

The large scale spiral structure of the Galactic magnetic field is disrupted locally
by turbulent motions of the gas. The amplitude of the perturbation will depend
on the turbulence energy density; on average data indicates that
the energy in the turbulent part of the magnetic field is of the same order
as the energy in the large scale field (see \cite{beck2001} and references therein).
In addition \cite{minter1996} showed that the power spectrum of the turbulent part of the
magnetic field follows a power law with an exponent of -5/3 at scales
smaller than 100~pc. At larger scales, the structure of the magnetic field
is dominated by the spiral pattern. 

The impact of this turbulent part of the magnetic field on polarization 
is to lower the polarization fraction with respect to what would be observed
with only the large scale magnetic field. This turbulent depolarization is stronger
at low Galactic latitudes as lines of sight in the Galactic plane 
cross more fluctuations than at high latitudes.
This effect is shown in Fig.~\ref{fig:turbul} where we have simulated the effect
of a turbulent magnetic field superposed on a large scale
field perpendicular to the line of sight ($B_x=B_y=1$~$\mu$G and $B_z=0$~$\mu$G).
The $x$, $y$ and $z$ components of the turbulent field are independent 1D Gaussian
random fields along the line of sight (10 000 points, cell size of 0.1~pc) 
with a power spectrum following a -5/3 power law at scales lower than 100~pc. 
For this exercise we assumed a constant cosmic ray density and $s=3$ (i.e $f_s=0.75$).
The polarization fraction is then computed by integrating numerically 
equations \ref{eq:I_sync}, \ref{eq:Q_sync} and \ref{eq:U_sync} for increasing values
of the depth $L$. This was repeated for increasing values of $B_{turb}$ from 0.01 to 10~$\mu$G.

As expected the polarization fraction decreases generally with the depth of the line
of sight and the amplitude of the turbulent motions. More and/or stronger turbulent
fluctuations reduce the coherence of the field and the polarization fraction.
In fact the behavior observed here depends solely on the ratio of $B_{turb}/B$ and not
on their respective absolute values.
We noted that for moderate values of $B_{turb}/B$, 
the polarization fraction reaches an asymptotic value. This values is reached
at larger line of sight depths for greater $B_{turb}/B$ values. 

The map of $g$ was computed by adding a turbulent vector field 
to the BSS magnetic field model for each sky coordinate. Therefore every lines of sight are
independent and no coherence on the plane of the sky due to turbulence is introduced.
The turbulent part added here allows
to set the proper average level of $g$ but it also modifies
the structure of $g$ as the depolarization due to turbulence depends on 
the length of the line of sight and the ratio of $B_\perp/B$. 
Nevertheless most of the structure of $g$ is set by the spiral structure of the field.

\begin{figure}
\includegraphics[width=\linewidth, draft=false]{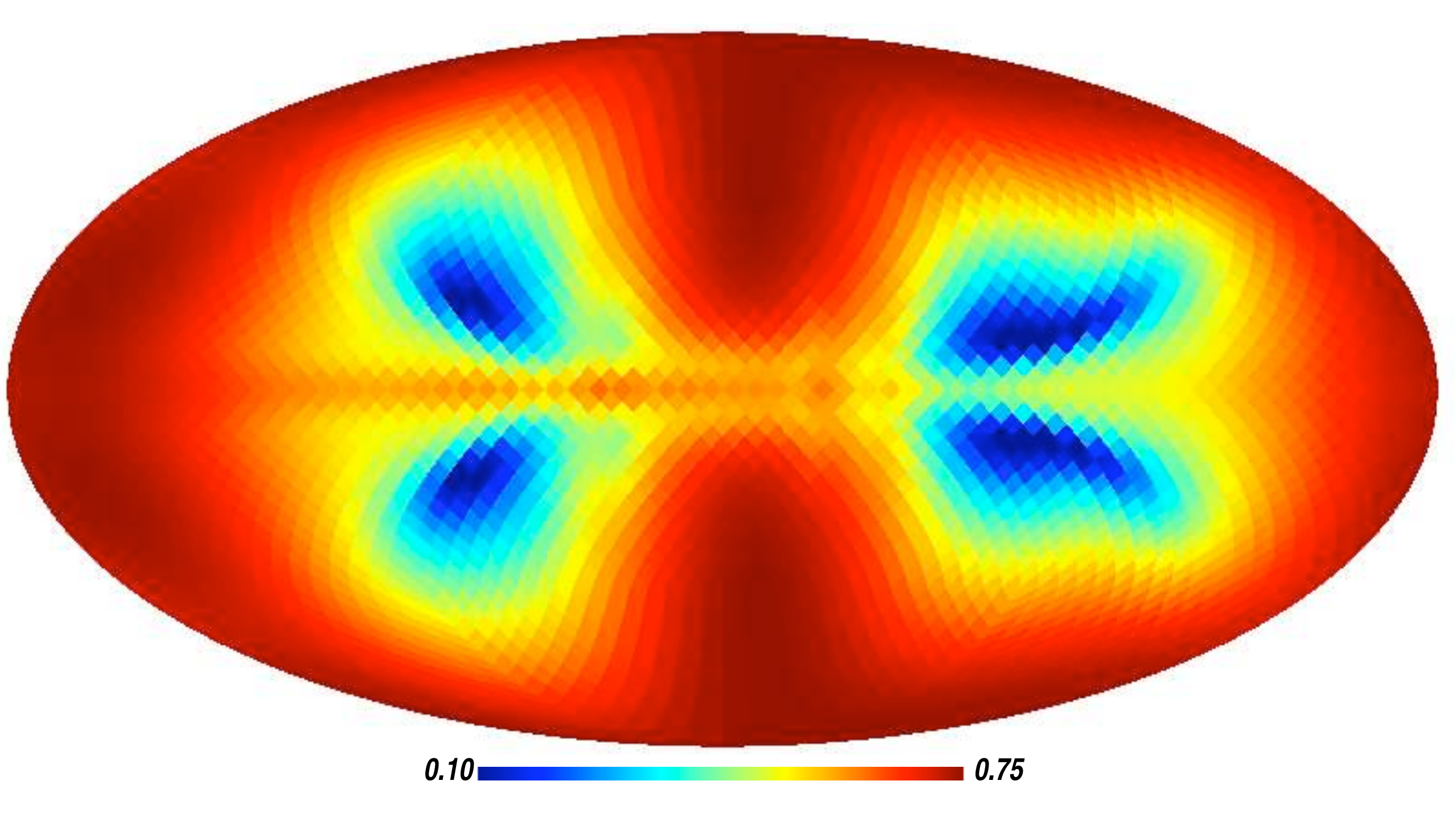}
\includegraphics[width=\linewidth, draft=false]{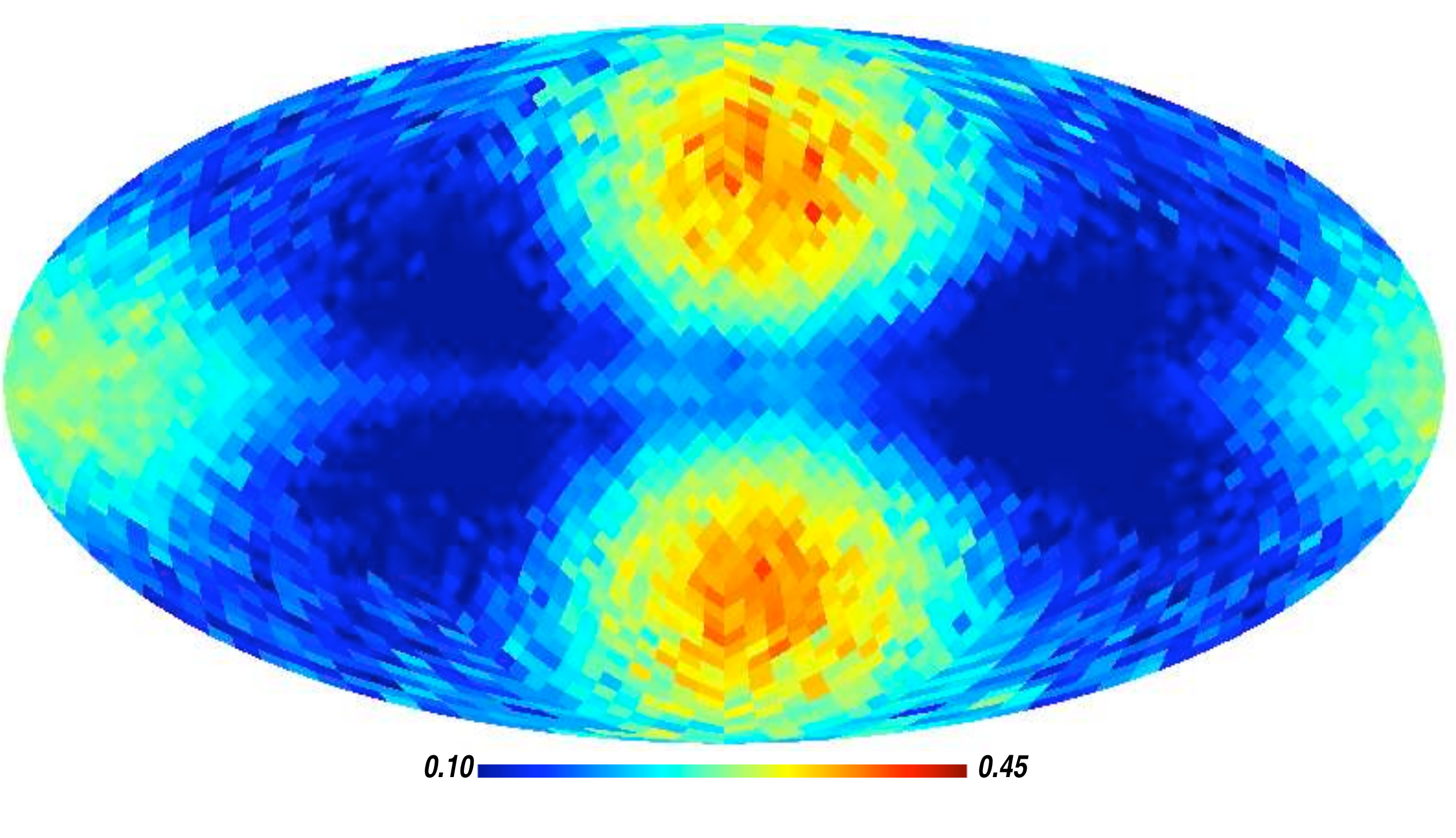}
\caption{\label{fig:g_model_turb} Model of the polarization fraction with (bottom) and without (top) turbulence.
Parameters of the magnetic field model are $p=-8.5^\circ$, $\chi_0=8^\circ$, 
$r_0=11$~kpc, $B_0= 3$~$\mu$G and turbulence of $\sigma_{turb}=1.7$~$\mu$G.}
\end{figure}

An example of a $gf_s$ map, with and without turbulence added is shown in Fig.~\ref{fig:g_model_turb}.
Depolarization due to turbulence is increased in the galactic plane (where line of sight are longer)
and in regions where $B_\perp/B$ is lower. The two bright spots above and under the Galactic center
corresponds to regions where the Galactic magnetic field is almost perpendicular to the 
line of sight.

\begin{figure}
\includegraphics[width=\linewidth, draft=false]{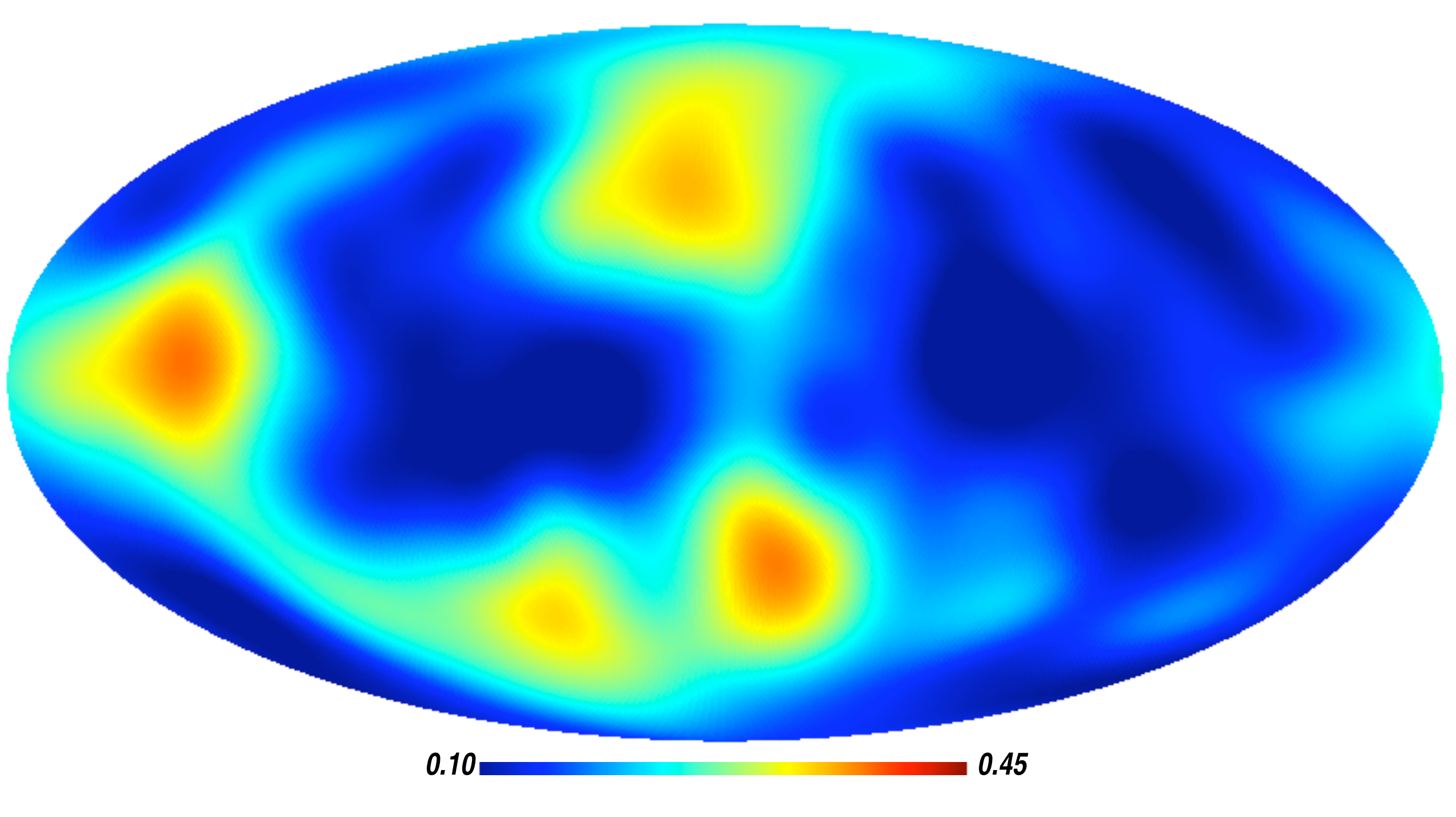}
\includegraphics[width=\linewidth, draft=false]{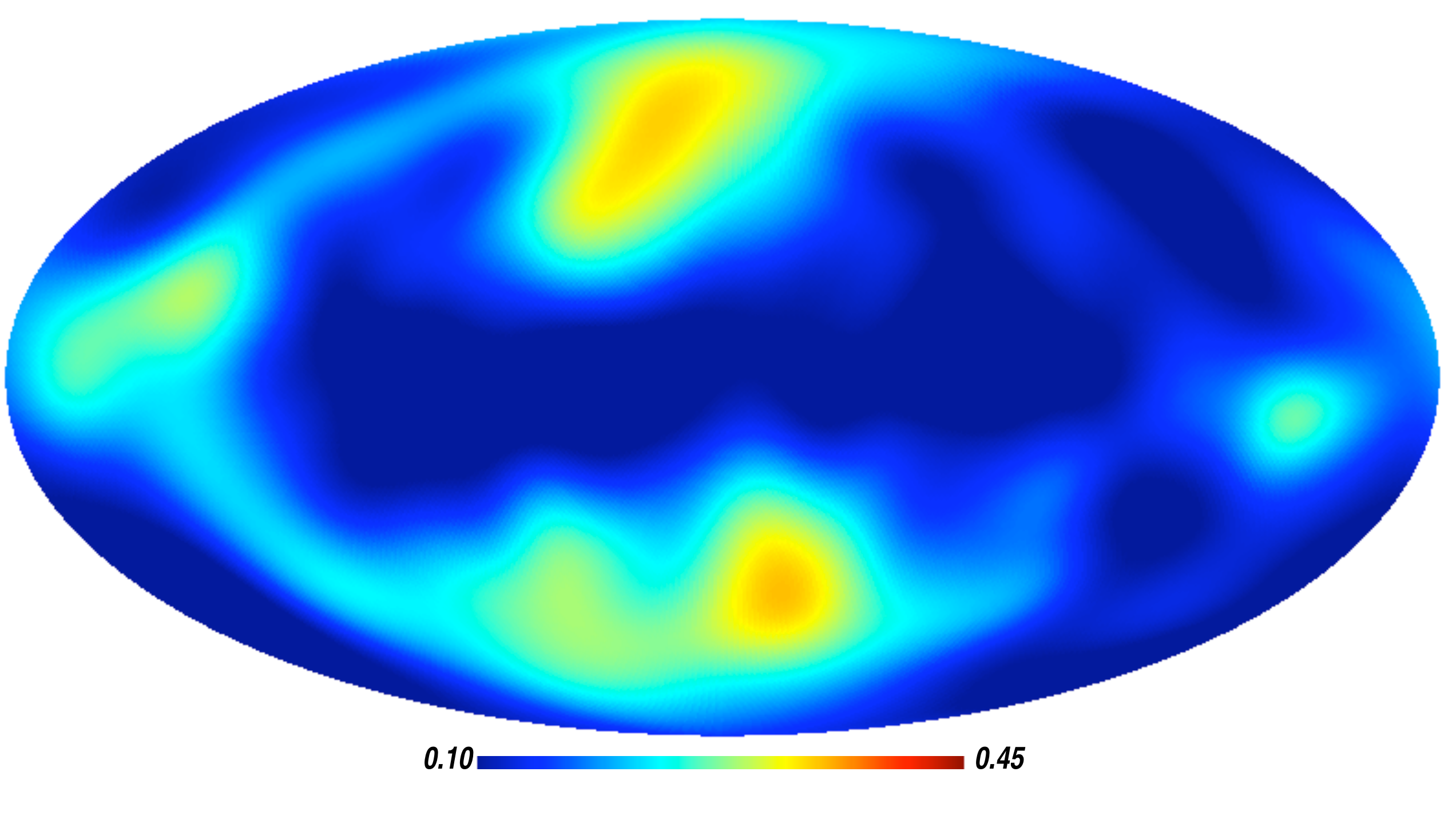}
\includegraphics[width=\linewidth, draft=false]{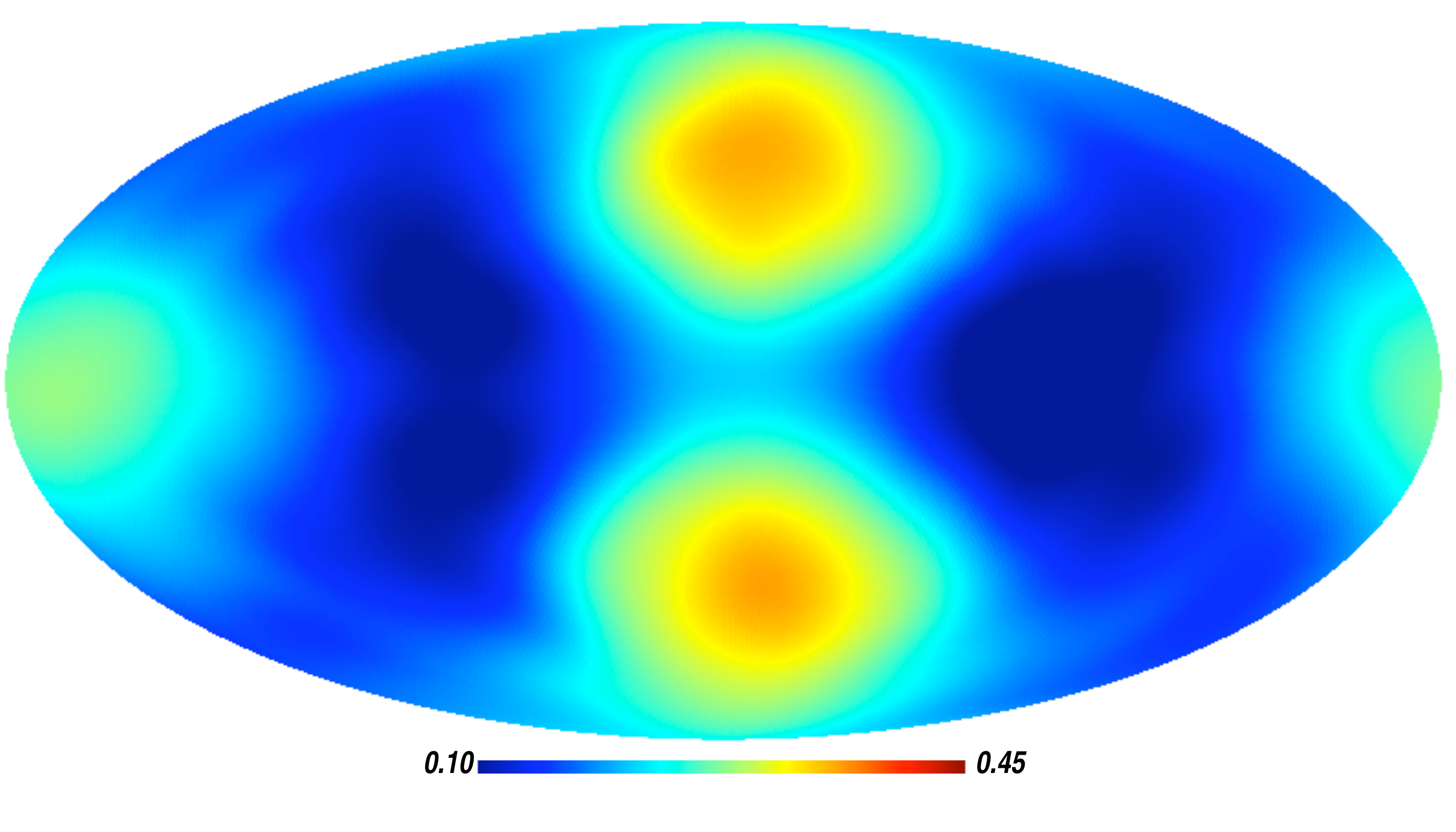}
\caption{\label{fig:g_model} Polarization fraction smoothed at $20^\circ$.
{\bf Top:} Model 2: constant $\beta_s=-3$. 
{\bf Middle:} Model 3: $\beta_s$ estimated with a anomalous emission proportional to $E(B-V)$. 
{\bf Bottom:} Model 4: $\beta_s$ estimated from polarization for $p=-8.5^\circ$, $\chi_0=8^\circ$, 
$r_0=11$~kpc, $B_0= 3$~$\mu$G and turbulence of $\sigma_{turb}=1.7$~$\mu$G.}
\end{figure}

\subsubsection{Fitting the magnetic field parameters}

Unlike \cite{page2007} we did not fit the parameters of the model on the polarization angle
which is very sensitive to high latitudes structures like Loop~1.
Instead the parameters are determined by fitting the polarization fraction map directly.
Our model is not depedant on the absolute values of $B_0$ and $B_{turb}$ but only to their ratio.
The parameters we are varying to fit the data are : $p$, $r_0$, $\chi_0$ and $B_{turb}$.
The fixed parameters are : $\gamma_{turb}=-5/3$, $l_{max}=100$~pc, $B_0=3$~$\mu$G and $n_e$.

The fitting of the parameters is done by minimizing the $\chi^2$ of the polarization fraction of
our Galactic magnetic field model
with respect to the polarization fraction deduced from models 2 and 3 and smoothed at $20^\circ$ 
(see Fig.~\ref{fig:g_model}).
Similar values are found for both model 2 and 3.
The best fit model gives $p=-8.5^\circ$, $\chi=8^\circ$, $r_0=11.0$~kpc, 
$B_{turb}/B_0 = 0.57$. The polarization fraction of our best fit model 
is shown in Fig.~\ref{fig:g_model}.
The correlation factor are 0.65 and 0.72 for model 2 and 3 respectively.

The parameters found for the large scale field are in very good 
agreement with what is obtained from the analysis of pulsar rotation measures 
and from a previous analysis of the WMAP data by \cite{jansson2007}.
In particular the pitch angle value we found ($p=-8.5^\circ$) is remarkably 
close to the values obtained using pulsar rotation measures 
\cite[]{han1994,han1999,indrani1999,beck2001,han2006b}
or polarized starlight absorption measurements
\cite[]{heiles1996} where values are in found in the range $-7< p< -11^\circ$.
On the other hand our result is not compatible with the result of \cite{page2007}
who obtained $p=-35^\circ$.

The values of $B_{turb}/B_0$ found for model 2 and 3 is also remarkably close to one obtained 
from analysis of the synchrotron emission \cite[]{phillipps1981,beck2001} which
gives $B_{turb}/B_0 \sim 0.66$, which is implies energy equipartition between the turbulent and large scale components.
Our result is also in agreement with the work of \cite{jones1992} who obtained $B_{turb}/B_0\sim=0.6$
using polarized extinction measurements at 2.2~$\mu$m.
On the other hand, and similarly to \cite{sun2007a} who did not consider the 
presence of anomalous emission at 23 GHz,
model 1 requires $B_{turb}/B_0 \sim 1.5$ to fit the data, which is significantly stronger
than the value obtained by \cite{phillipps1981,beck2001}.
This is due to the fact that anomalous emission contributes a significant fraction of
the intensity at 23~GHz and that it is not polarized. 
If (unpolarized) anomalous emission is not taken into account at 23~GHz, 
a stronger depolarization (and therefore more turbulence) is needed 
to account for the low polarization fraction. The incapacity of model~1 to reproduce the right level
of $B_{turb}/B_0$ is a strong evidence for the presence of an unpolarized anomalous emission at 23~GHz. 
Therefore in the following we will not consider model~1 anymore.

\begin{figure}
\includegraphics[width=\linewidth, draft=false]{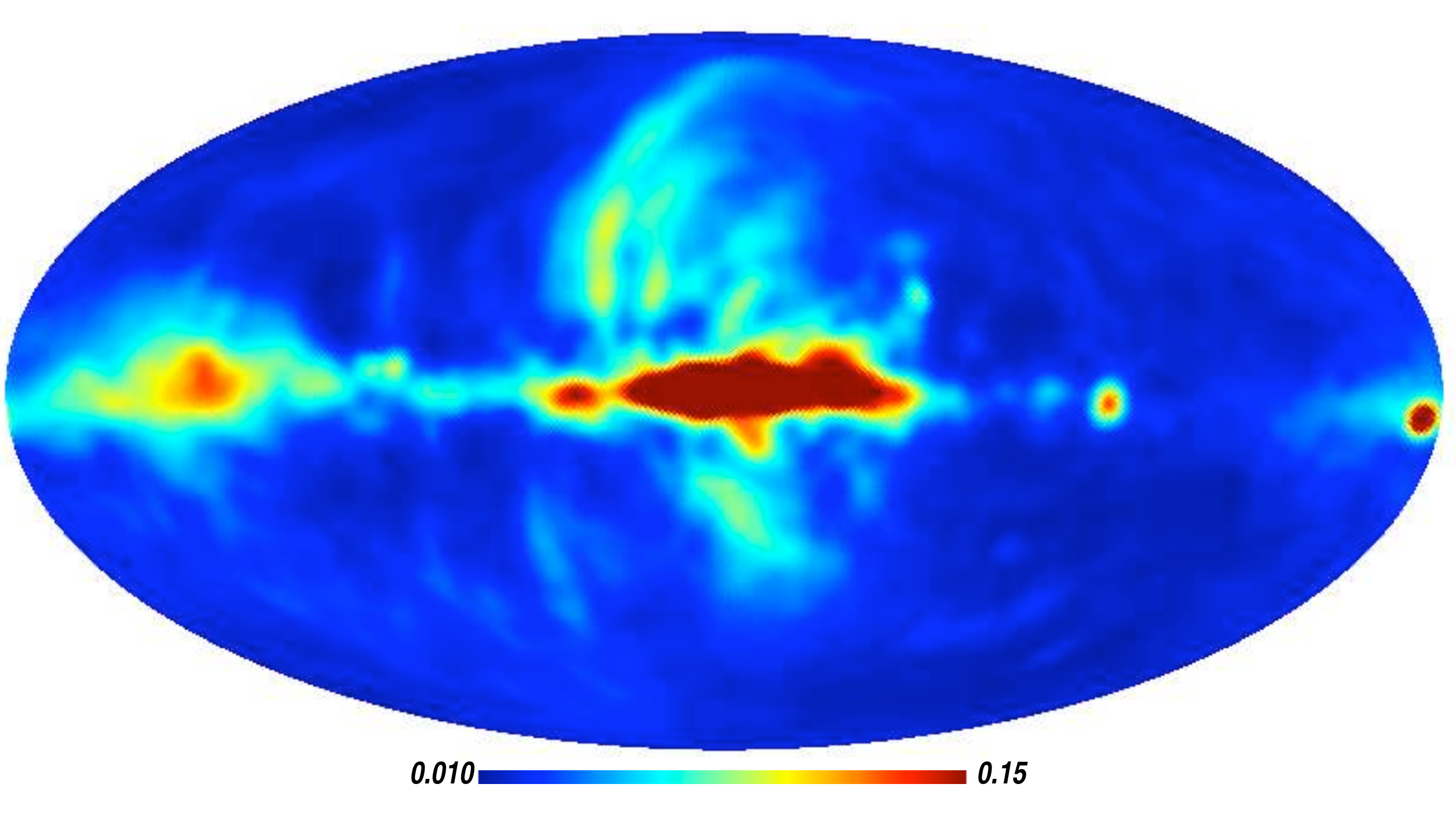}
\includegraphics[width=\linewidth, draft=false]{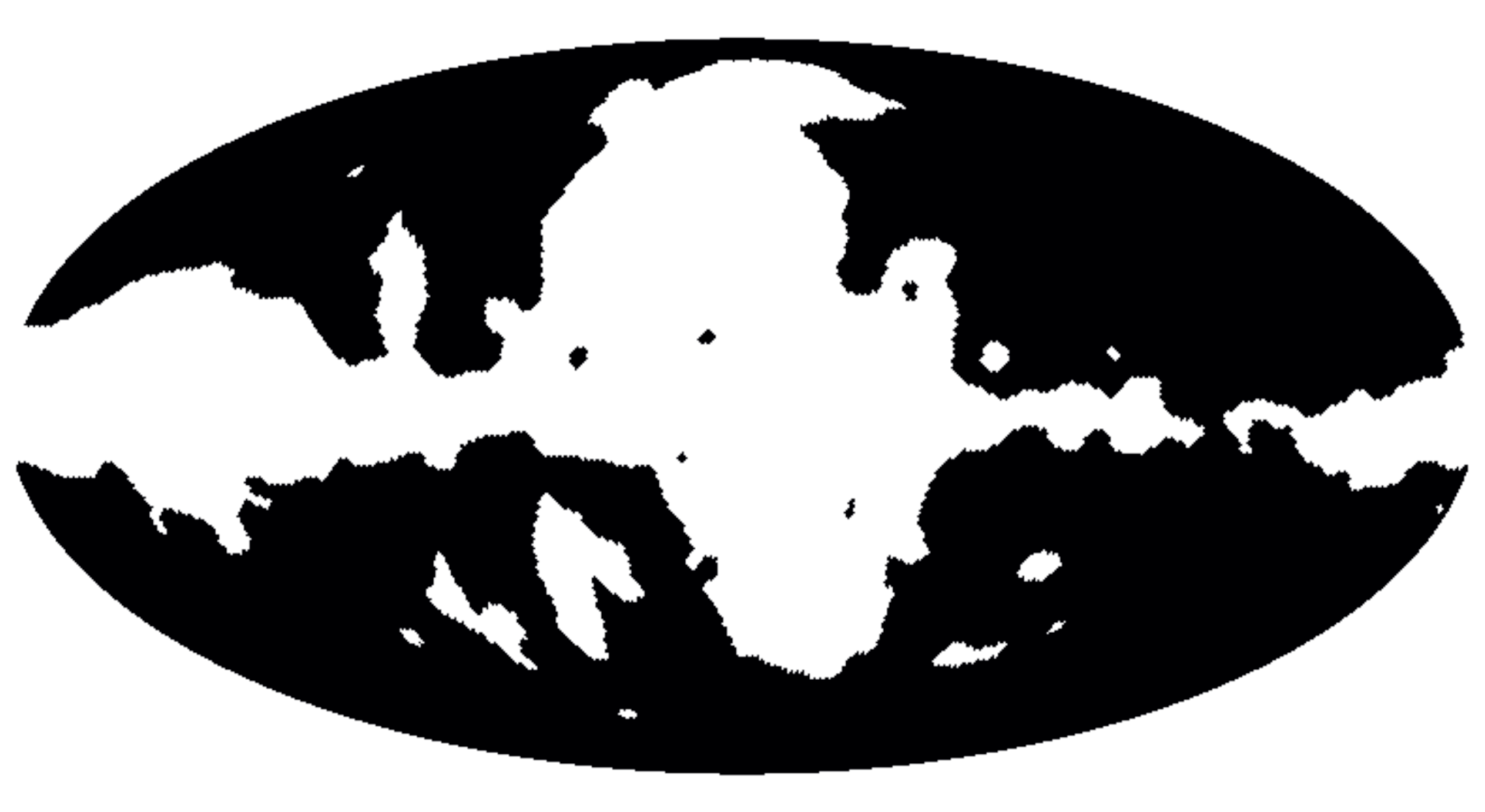}
\caption{\label{fig:polar23} {\bf Top:} Polarized intensity (in mK) at 23 GHz, smoothed at $5^\circ$.
{\bf Bottom:} Mask showing regions of signal-to-noise ratio greater than 3.}
\end{figure}

\subsection{Model of synchrotron emission}

We build the synchrotron spectral index using 
\begin{equation}
\label{eq:beta_s_polar}
\beta_s = \frac{\log(M_{23}/S_{408}) }{\log(23/0.408)} 
\end{equation}
where $M_{23}$ is the model of the intensity at 23~GHz obtained from polarization:
\begin{equation}
M_{23} = \frac{P_{23}}{g  f_s}.
\end{equation}
$P_{23}$
is the polarization map from WMAP smoothed at 5$^\circ$ (shown in Fig.~\ref{fig:polar23}),
$f_s$ is given by Eq.~\ref{eq:fs} and the depolarization fraction $g$ is given by our model, 
smoothed at $20^\circ$ to remove any small scale fluctuations introduced by the turbulent part
of the magnetic field.

A reliable estimate of $M_{23}$ can only be made for pixels where the signal-to-noise
ratio of the polarization data is significant. We defined a mask shown in Fig.~\ref{fig:polar23}
which gives regions with polarization signal-to-noise ratio large enough to estimate the synchrotron
intensity. Inside this mask $\beta_s$ was estimated using Eq.~\ref{eq:beta_s_polar}.
Outside the polarization mask, E(B-V) is mostly lower than 1
which corresponds to regions of the sky where the anomalous emission correlates well
with extinction. Therefore we used the estimate of $\beta_s$ given by model 3 in regions outside
the polarization mask.
The resulting $\beta_s$ map at a resolution of $5^\circ$ is shown in the bottom 
panel of Fig.~\ref{fig:beta}.
Even if $\beta_s$ was estimated with two different models, we note that there are no striking
discontinuities at the boundaries of the mask.
Compared to models 1 and 3 the galactic plane does not stand out in this spectral index map. 
In that respect the spectral index map deduced from the polarization data is much more similar to the 
ones determined between 408, 1420 and 2326 MHz by \cite{giardino2002} and by \cite{platania2003}.

The histogram of $\beta_s$ is shown in Fig.~\ref{fig:histo_beta} (dashed line).
It has an average value of -3.00 and a standard deviation of 0.06. The minimum dispersion
of $\beta_s$ is thus obtained for model 4.

\begin{figure}
\includegraphics[width=\linewidth, draft=false]{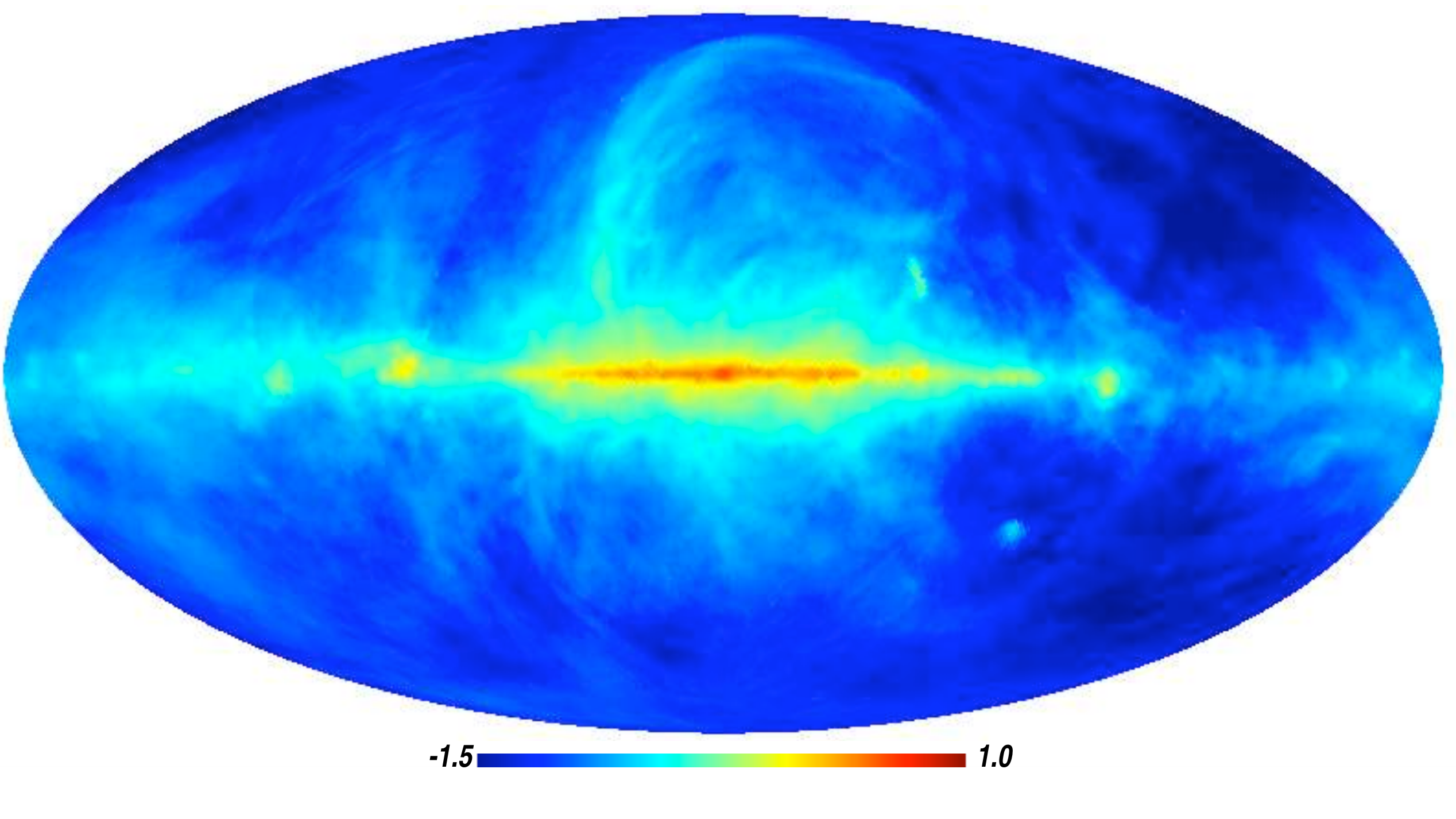}
\includegraphics[width=\linewidth, draft=false]{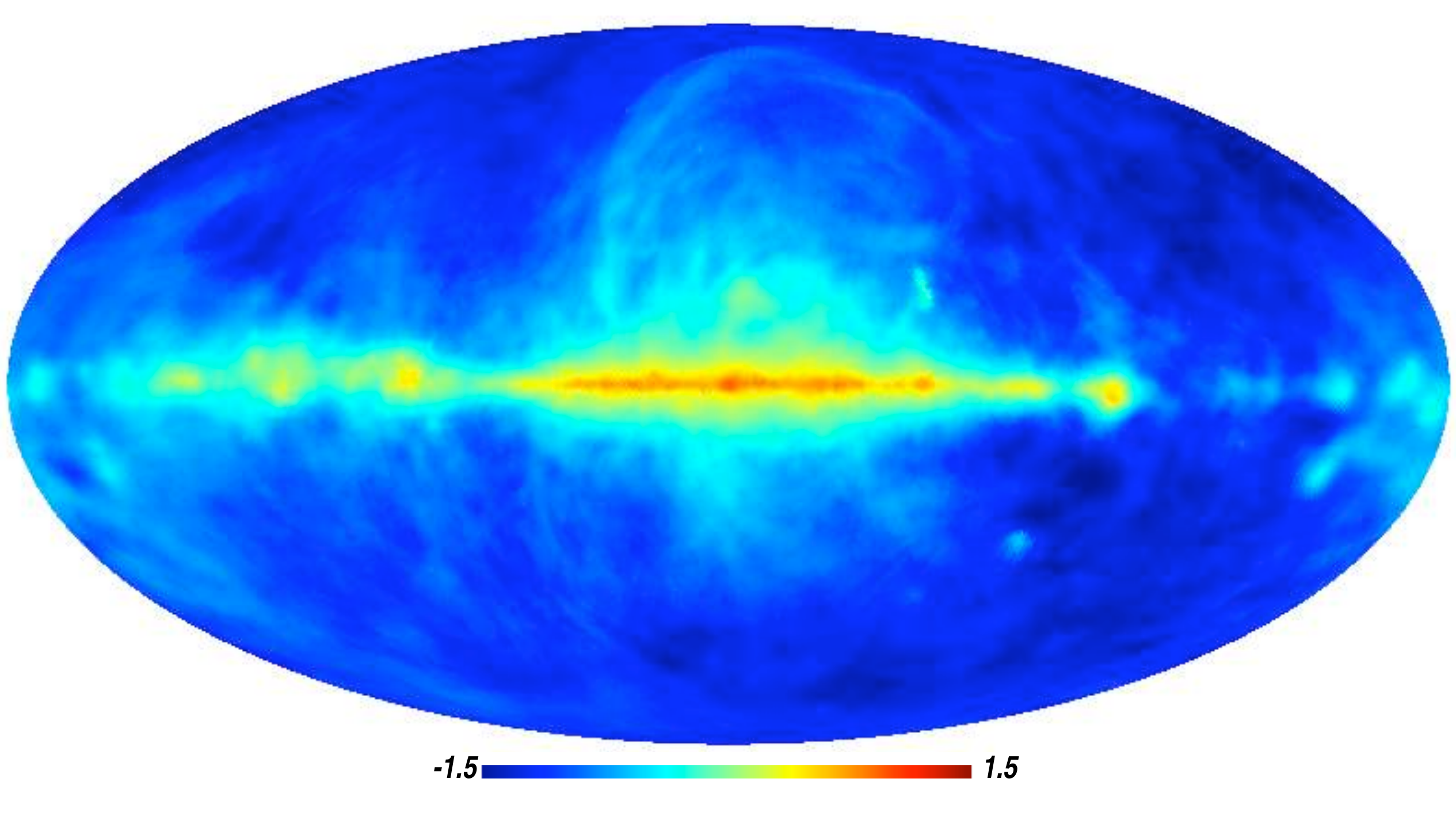}
\includegraphics[width=\linewidth, draft=false]{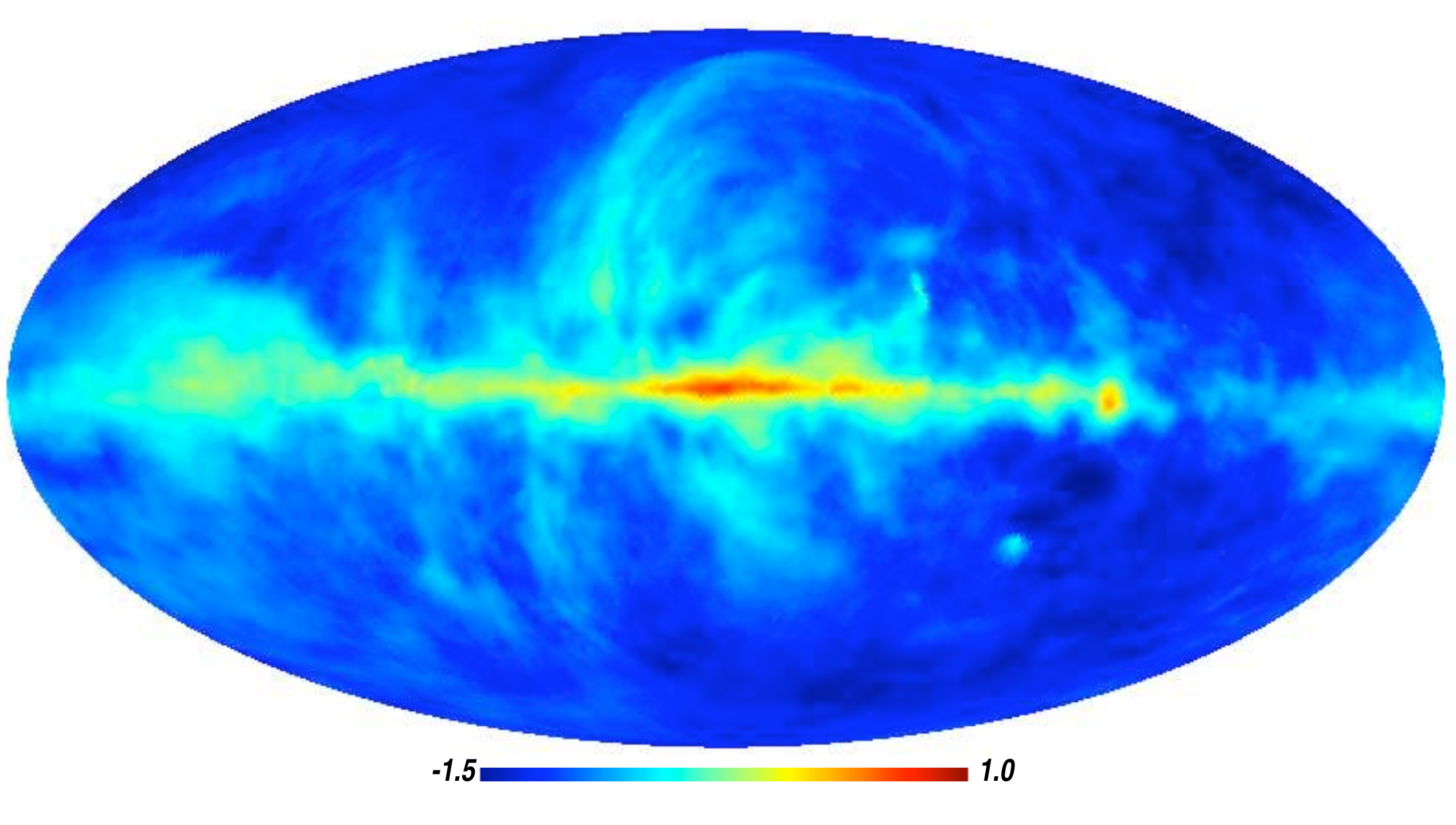}
\caption{\label{fig:synchrotron} Synchrotron emission (in log$_{10}$(mK)) at 23~GHz for models 2 (top), 
3 (middle) 
and 4 (bottom). The synchrotron map for model 2 (top) is a scaling of the 408~MHz map with
a constant $\beta_s=-3$.}
\end{figure}

\begin{figure}
\includegraphics[width=\linewidth, draft=false]{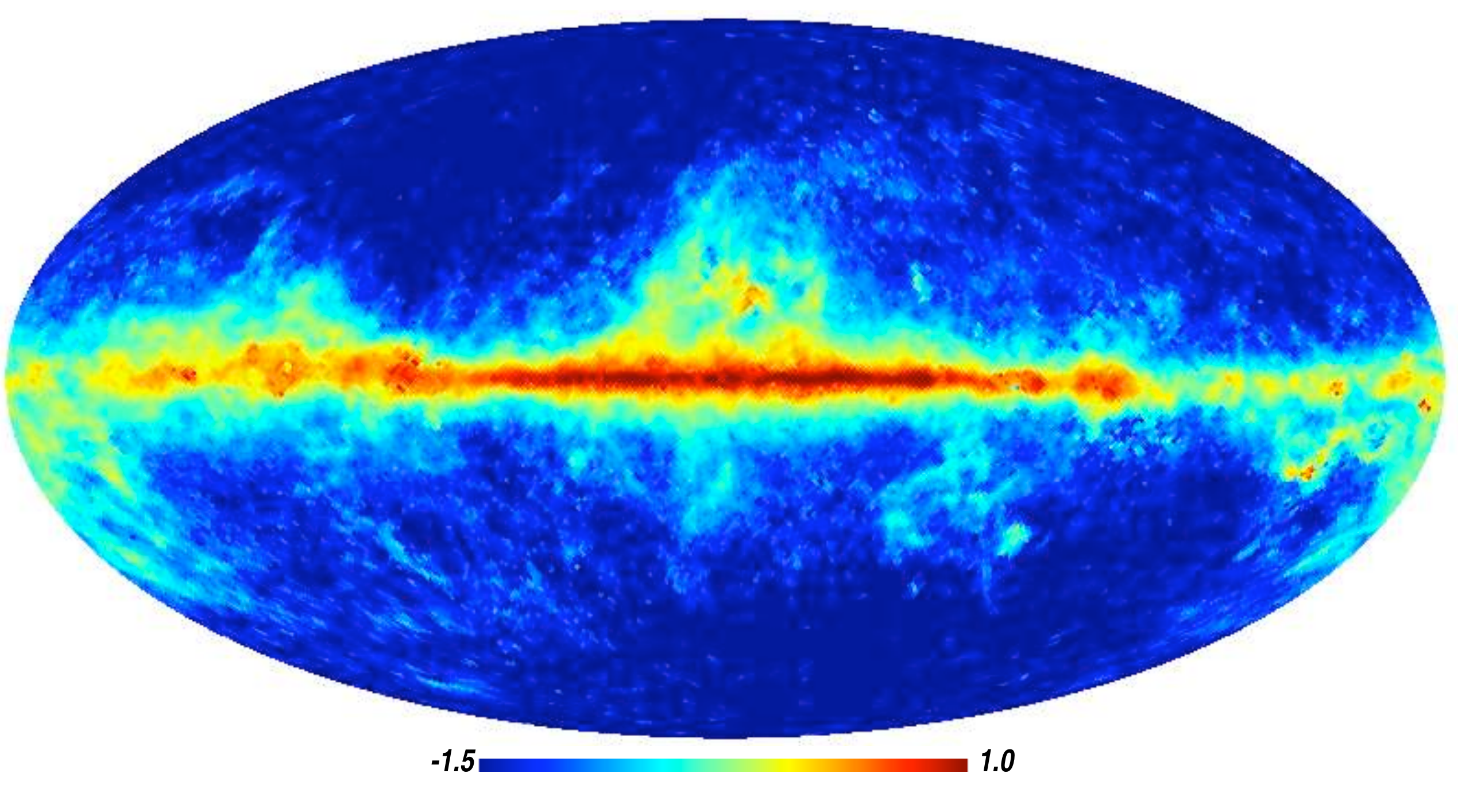}
\includegraphics[width=\linewidth, draft=false]{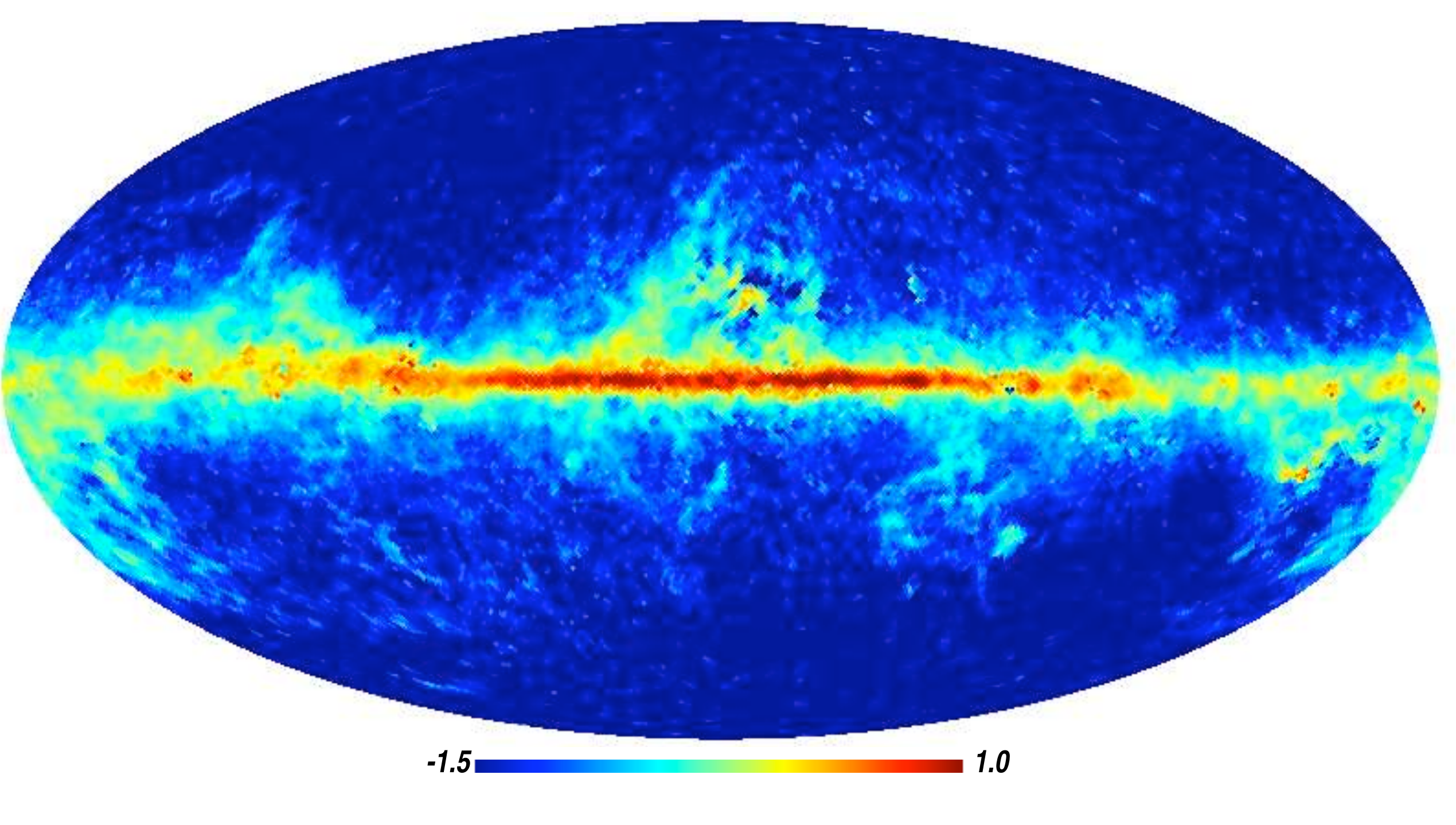}
\includegraphics[width=\linewidth, draft=false]{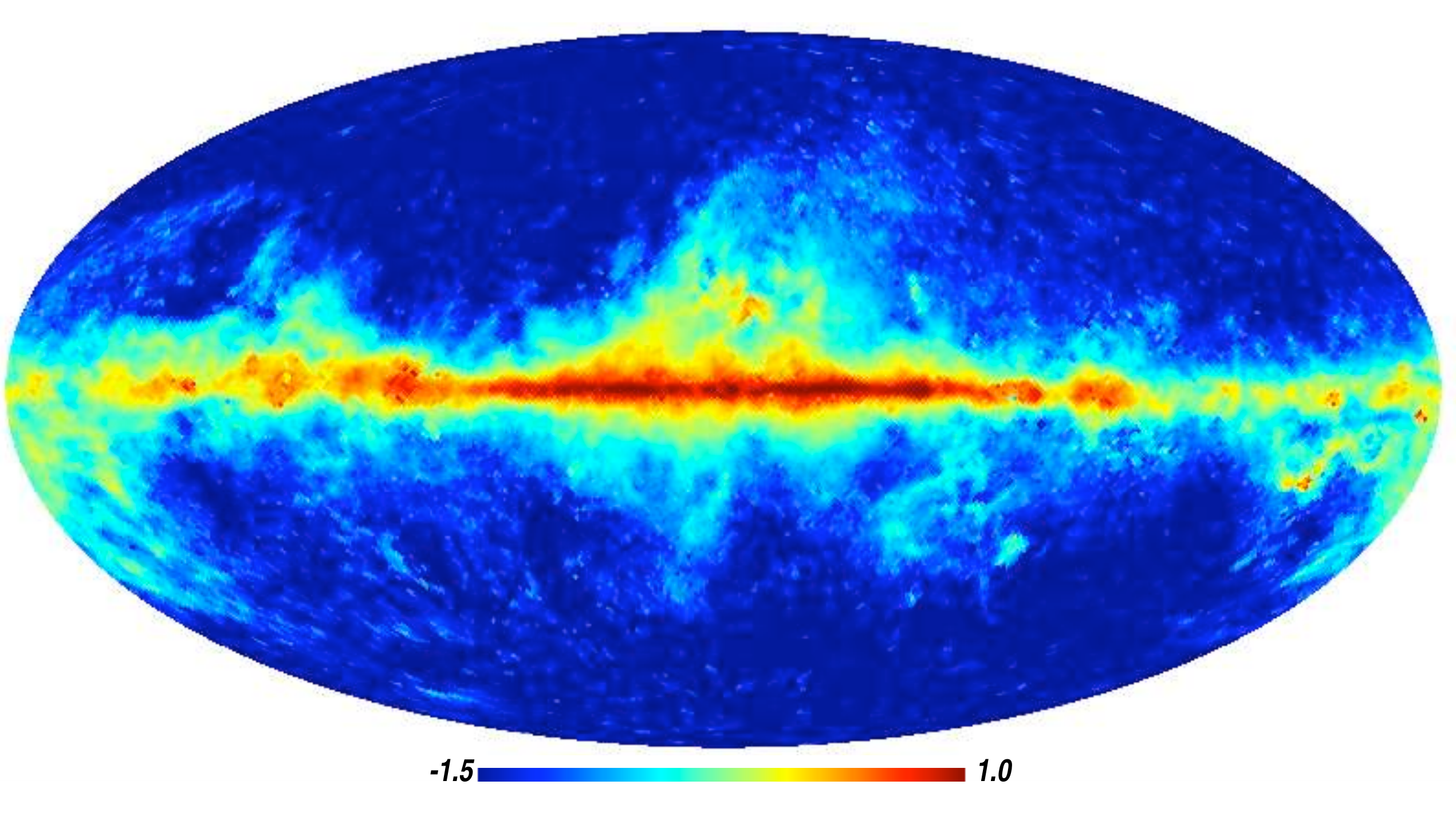}
\caption{\label{fig:spinning} Estimate of the anomalous emission (in log$_{10}$(mK)) at 23~GHz
for models 2 (top), 3 (middle) and 4 (bottom).}
\end{figure}

\section{Anomalous emission in the WMAP data}

\label{sec:anomalous}

The predicted synchrotron intensity at 23~GHz for models 2, 3 and 4 is shown 
in Fig.~\ref{fig:synchrotron}. 
For model 3 (Fig.~\ref{fig:synchrotron}-middle), there is more synchrotron emission in the plane
than for the two other models. This is due to the spectral index used which is significantly
shallower in the plane (see Fig \ref{fig:beta}-middle). The synchrotron emission for models 2 and 4
are very similar in overall intensity but there are significant spatial variations, especially
in the Galactic plane, partly due to the fact that model~4 does not include spatial variations
of $g$ at scales smaller than 20$^\circ$.

The residual emission, when free-free and synchrotron are removed from the 23 GHz
data, are shown in Fig.~\ref{fig:spinning} for models 2, 3 and 4. 
We attribute this residual to the anomalous emission. 
The anomalous emission obtained for models 2 (constant $\beta_s=-3$) and model 4 ($\beta_s$
obtained from polarization) are very similar. The difference between these two models
is well fitted by a Gaussian of average 0.002~mk and $\sigma=0.015$~mK.
For both models, the correlation factor of the anomalous emission with extinction (E(B-V)) is 0.94.
This analysis shows that the synchrotron emission at 23~GHz can
be rather well estimated, to first approximation, with an extrapolation of the 408~MHz with 
a constant spectral index $\beta_s=-3$. In the following we describe the properties of the anomalous
emission for model~4.

\begin{figure}
\includegraphics[width=\linewidth, draft=false]{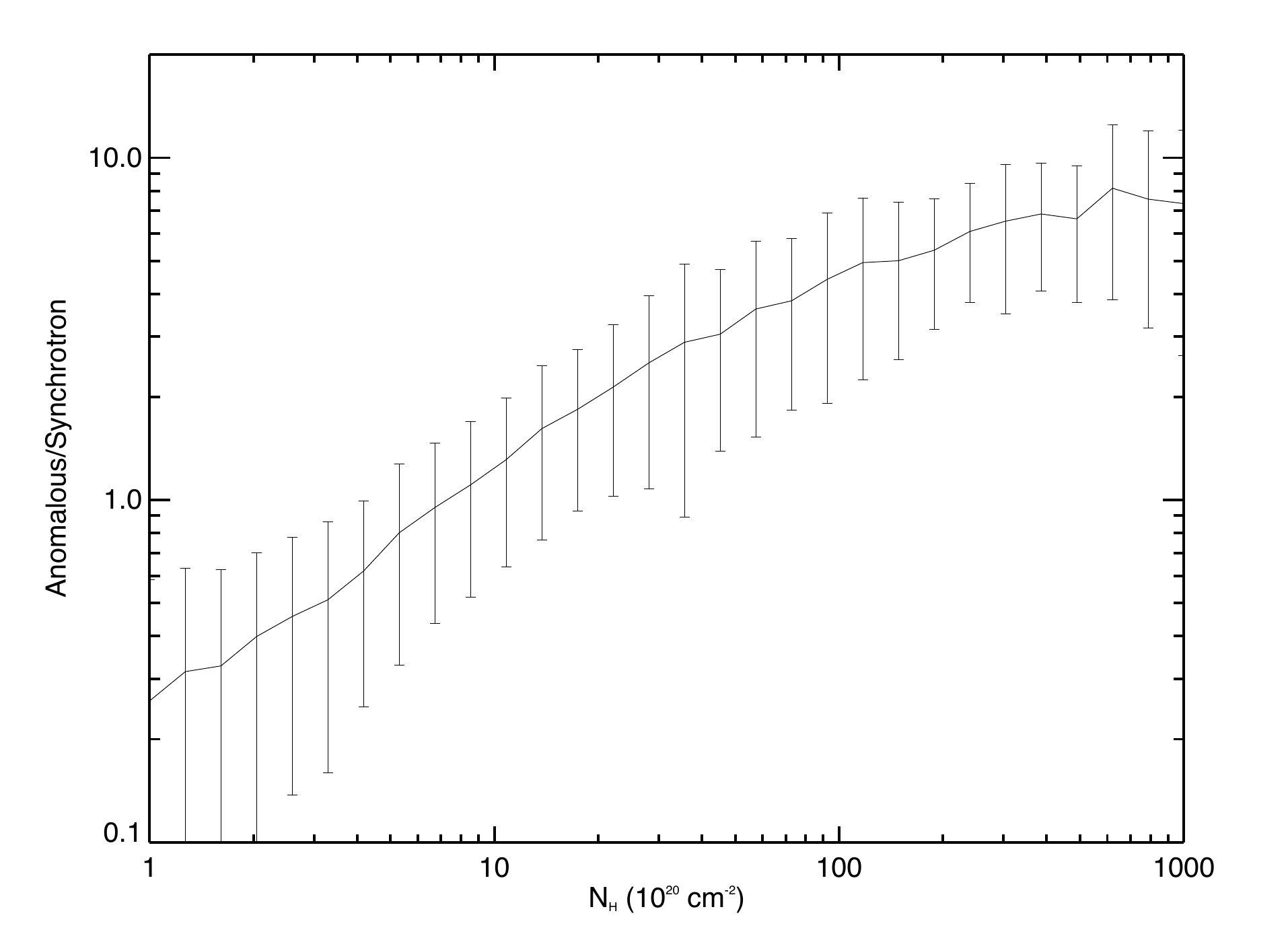}
\caption{\label{fig:anomalous_synchrotron_ratio} Ratio of anomalous emission to synchrotron at 23 GHz
(model~4) as a function of column density. In the plane, the anomalous is on average several times
higher than the synchrotron emission. In the most diffuse regions at high Galactic latitude 
the synchrotron emission overcomes the anomalous emission in total intensity. 
The error bars represent the standard deviation of the anomalous to synchrotron ratio
in each bin of column density.}
\end{figure}

The anomalous emission is dominating the signal from the Galactic plane to mid-latitude regions, 
with a mean intensity up to ten times higher than synchrotron 
(see Fig. \ref{fig:anomalous_synchrotron_ratio}). 
On the other hand, we note that the average level of synchrotron emission at high latitude
is about three times that of the anomalous emission, likely due to the large scale height
of cosmic ray electrons.
The Gould belt's cloud have strong anomalous emission and known high-latitude cirrus regions 
like the North Celestial Loop are also seen in anomalous emission.

\begin{figure}
\includegraphics[width=\linewidth, draft=false]{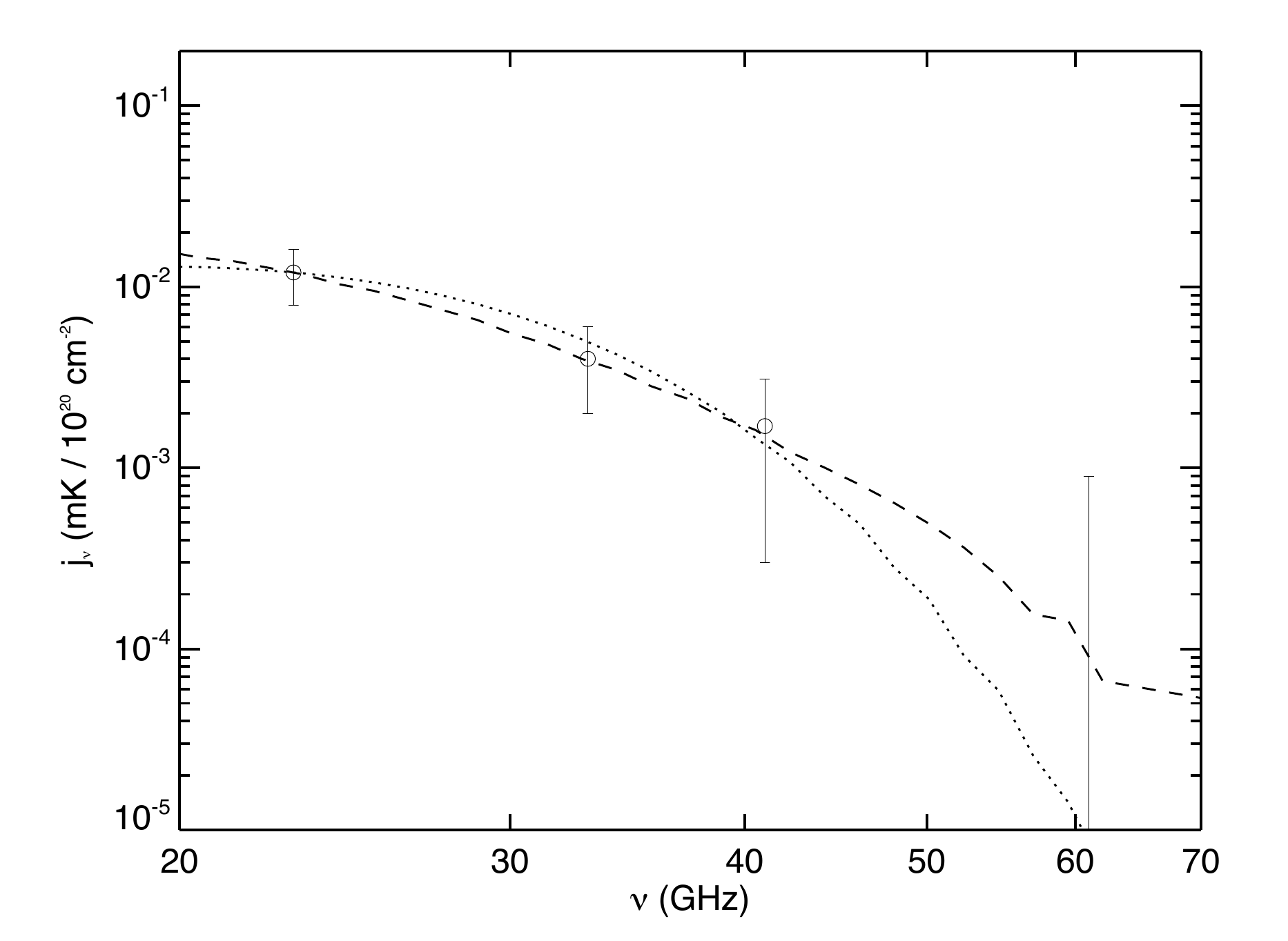}
\caption{\label{fig:anomalous_spectrum} Spectrum of the anomalous emissivity obtained
by correlating the anomalous from model~4 with $N_H = 8.0\times 10^{21} \, E(B-V)$. 
The correlation was estimated for $N_H>10^{21}$~cm$^{-2}$. 
The uncertainty is the standard deviation of the emissivity observed for the selected pixels. 
The emissivity at 61~GHz is compatible with zero.
The curves are the emissivity spectra, normalized at 23~GHz, 
given by \cite{draine1998a} for the WNM (dashed) and by 
\cite{ysard2008} for PAHs ranging from 50 to 96 carbon atoms (dotted).}
\end{figure}

The anomalous emission was also extracted at 33, 41 and 61~GHz using the 
synchrotron spectral index map of model~4.
The emissivity per hydrogen atom at each frequency of the anomalous emission
was estimated by correlating it with $N_H$~(cm$^{-2}$)~$ = 5.8\times 10^{21} E(B-V)$ 
(see Fig~\ref{fig:anomalous_spectrum}) for $N_H>10^{21}$~cm$^{-2}$.
The spectrum obtained using model~4 is shown in Fig.~\ref{fig:anomalous_spectrum}; 
the anomalous emission spectrum is the same within 1\% for models~2 and 4.
In Fig.~\ref{fig:anomalous_spectrum} the anomalous emission spectrum is compared to
the spinning dust model of \cite{draine1998a} for the Warm Neutral Medium (WNM)
and with the model of \cite{ysard2008} (for PAH sizes ranging from 50 to 96 carbon atoms).
Both models reproduce well the observed spectrum. A detailed analysis of the 
anomalous emission spectrum is left to a further study.

\section{Conclusion}

\label{sec:conclusion}

In this paper we presented a new analysis of the Galactic emission at 23 GHz as observed by WMAP.
The goal of this study was to combine the total intensity and polarization 
data from WMAP to separate the synchrotron emission and the anomalous emission at this frequency.
Our analysis is based on the hypothesis that 
1) the 23 GHz intensity data of WMAP is dominated by CMB, synchrotron, free-free and anomalous emission, 
2) the 23 GHz polarization data is dominated by synchrotron only
3) the average spectral index of synchrotron between 408 MHz and 23 GHz is close to $\beta_s=-3$,
and 4) the Galactic magnetic field has a BSS large scale structure with a turbulent part 
that follows a -5/3 power spectrum  at scales smaller than 100 pc.

We used the WMAP 23~GHz polarization data to constrain the parameters of the magnetic field.
The data are found compatible with a pitch angle of $-8.5^\circ$ 
and a ratio of the turbulent to regular part of the magnetic field of
$B_{turb}/B_0=0.57$. The parameters of our Galactic magnetic field model
are in good agreement with what is found with pulsar rotation measures, radio synchrotron emission
and polarized extinction in the near-infrared.

Based on our Galactic magnetic field model and the WMAP polarization data, we produced
a map of the synchrotron spectral index in total intensity, between 408 MHz and 23 GHz.
The map of $\beta_s$ we found does not show a strong Galactic plane feature,
in accordance with what was found in the radio by \cite{giardino2002,platania2003}. 
Extrapolating the Haslam 408 MHz with this
new estimate of the synchrotron spectral index, we could separate the contribution
of the synchrotron and anomalous emission at 23, 33, 41 and 61~GHz.

The all-sky anomalous emission at 23 GHz is found to be correlated to E(B-V) (or to $A_v$)
with a correlation factor of 94\%. 
It is well detected in the Galactic plane, in Gould Belt's clouds and in the main high-latitude 
cirrus clouds.
In the Galactic plane the anomalous emission is up to ten times stronger than the synchrotron
but it is only one third of the synchrotron in the most diffuse regions of the sky 
where the synchrotron, while very smooth, has an average level stronger than the anomalous emission, 
possibly due to the larger scale height of cosmic ray electrons compare to interstellar dust.
Under the hypothesis that the synchrotron spectral index does not vary with frequency over the WMAP range, 
we found that the anomalous emission has a spectrum from 23 to 61 GHz in accordance with the model of spinning dust
from \cite{draine1998a}. A deeper analysis of the origin of the anomalous emission and
of its consequences regarding the properties of interstellar dust will be the subject of a forthcoming paper.

The presence in the 20-200 GHz range of several diffuse Galactic components with similar intensities and
some spatial correlation makes the extraction of the CMB a complex task. In order to achieve sufficient accuracy
on the cosmological signal the component separation needs to take advantage of the knowledge on 
the properties of diffuse Galactic emission. 
In this paper we have shown that combining temperature and polarization data helps to improve
our model of Galactic emissions. The higher signal-to-noise ratio, the higher angular resolution,
the larger frequency coverage and the increase number of frequency bands of the Planck data 
will also certainly help to make progress in this area.

\vspace*{0.2cm}

Some of the results in this paper have been obtained using the HEALPix package
\cite[]{gorski2005}. This work was partly supported by the Canadian Space Agency.

\end{document}